\documentclass[10pt,journal,compsoc]{IEEEtran}
\ifCLASSOPTIONcompsoc
  \usepackage[nocompress]{cite}
\else
  \usepackage{cite}
\fi
\ifCLASSINFOpdf
\else
\fi
\usepackage{times}  
\usepackage{helvet} 
\usepackage{courier}  
\usepackage[hyphens]{url}  
\usepackage{graphicx} 
\usepackage{amsmath}
\usepackage{caption} 
\usepackage{float}
\usepackage{algorithm}  
\usepackage{algorithmicx}
\usepackage{algpseudocode} 
\usepackage{multirow}
\usepackage{color}
\usepackage{multicol}

\begin{document}
%
\title{Towards Improving Embedding Based Models of Social Network Alignment via Pseudo Anchors}
%
%
%
%

\author{Zihan~Yan,~
        Li~Liu,~
        Xin~Li,~\IEEEmembership{Member,~IEEE},
        William K.~Cheung,~\IEEEmembership{Member,~IEEE},
        Youmin~Zhang,~
        Qun~Liu,~
        and ~Guoyin~Wang,~\IEEEmembership{Senior Member,~IEEE}
\IEEEcompsocitemizethanks{\IEEEcompsocthanksitem Ziyan Yan, Li Liu, Youmin Zhang, Qun Liu and Guoyin Wang are with the Chongqing Key Laboratory of
Computational Intelligence, Chongqing University of Posts and Telecommunications, Chongqing 400065, China.\protect\\
Corresponding author: Li Liu. E-mail: liliu@cqupt.edu.cn
\IEEEcompsocthanksitem Xin Li is with Beijing Institute of Technology, Beijing, China, 100081.
\IEEEcompsocthanksitem Li Liu and William K. Cheung are with Hong Kong Baptist University, Hong Kong, China.}
\thanks{Manuscript received }}

%
%

\markboth{Journal of \LaTeX\ Class Files,~Vol.~14, No.~8, August~2015}%
{Shell \MakeLowercase{\textit{et al.}}: Bare Demo of IEEEtran.cls for Computer Society Journals}
%



\IEEEtitleabstractindextext{%
\begin{abstract}
Social network alignment aims at aligning person identities across social networks.
Embedding based models have been shown effective 
for the alignment where the structural proximity preserving objective is typically adopted for the model training. With the observation that ``overly-close'' user embeddings are unavoidable for such models causing alignment inaccuracy, 
we propose a novel learning framework which tries to enforce the resulting embeddings to be more widely apart among the users via the introduction of carefully implanted pseudo anchors.
We further proposed a meta-learning algorithm to guide the updating of the pseudo anchor embeddings during the learning process. The proposed intervention via the use of pseudo anchors and meta-learning allows the learning framework to be applicable to a wide spectrum of network alignment methods.
We have incorporated the proposed learning framework into several state-of-the-art models. Our experimental results demonstrate its efficacy where the methods with the pseudo anchors implanted can outperform their counterparts without pseudo anchors by a fairly large margin, 
especially when there only exist very few labeled anchors.

\end{abstract}

}

\maketitle

\IEEEdisplaynontitleabstractindextext

%
\IEEEpeerreviewmaketitle

\IEEEraisesectionheading{\section{Introduction}\label{sec:introduction}}

%
%
%
%
\IEEEPARstart{O}{nline} social networking is pervasive, with an enormous number of people communicating via multiple platforms. Social network alignment aims at aligning user accounts across such networks without knowing the real user identity. This alignment task has attracted widespread attention in both industry and academia due to its significant impact in many applications, e.g., user behavior prediction~\cite{DBLP:conf/aaai/JiangCYXY16}, friend recommendation~\cite{DBLP:journals/sigkdd/ShuWTZL16}, and identity verification~\cite{DBLP:conf/www/GogaLPFST13}.

Embedding based methods have been found effective for social network alignment where each network node is embedded into a low-dimensional space which in turn can support subsequent downstream prediction \cite{DBLP:journals/access/LiuZFZHZ19,DBLP:conf/ijcai/LiuCLL16,DBLP:conf/infocom/0002LZTWZ18,DBLP:conf/www/ChuFYZHB19,DBLP:journals/tkde/LiuLCL20,DBLP:conf/kdd/ZhangT16}. The key idea is to find a mapping of users across social networks under the supervision of labeled users' accounts (so-called anchor users). Compared to the earlier works \cite{DBLP:conf/aaai/TanGCQBC14,DBLP:conf/ijcai/ZhangY15}, the use of the embedding approach can avoid the computation of matrix inverse and is more suitable for alignment tasks over large-scale networks. Nevertheless, the structural proximity preserving objective typically adopted by the embedding algorithm can result in ``overly-close'' embeddings for nodes in a dense neighborhood structure. That makes it harder to differentiate from each other in the embedding space, and thus harder to align users across social networks. This limitation in principle can be alleviated by leveraging observed anchor pairs across the networks via (semi-)supervised learning. However, compared to the entire volume of social network users, only a small portion of users can be identified as the ground truth anchors across multiple social network platforms.
Lacking enough anchor pairs will result in degradation of the embedding based alignment methods.

\begin{figure*}[htbp]
\centering
\includegraphics[width=1\textwidth]{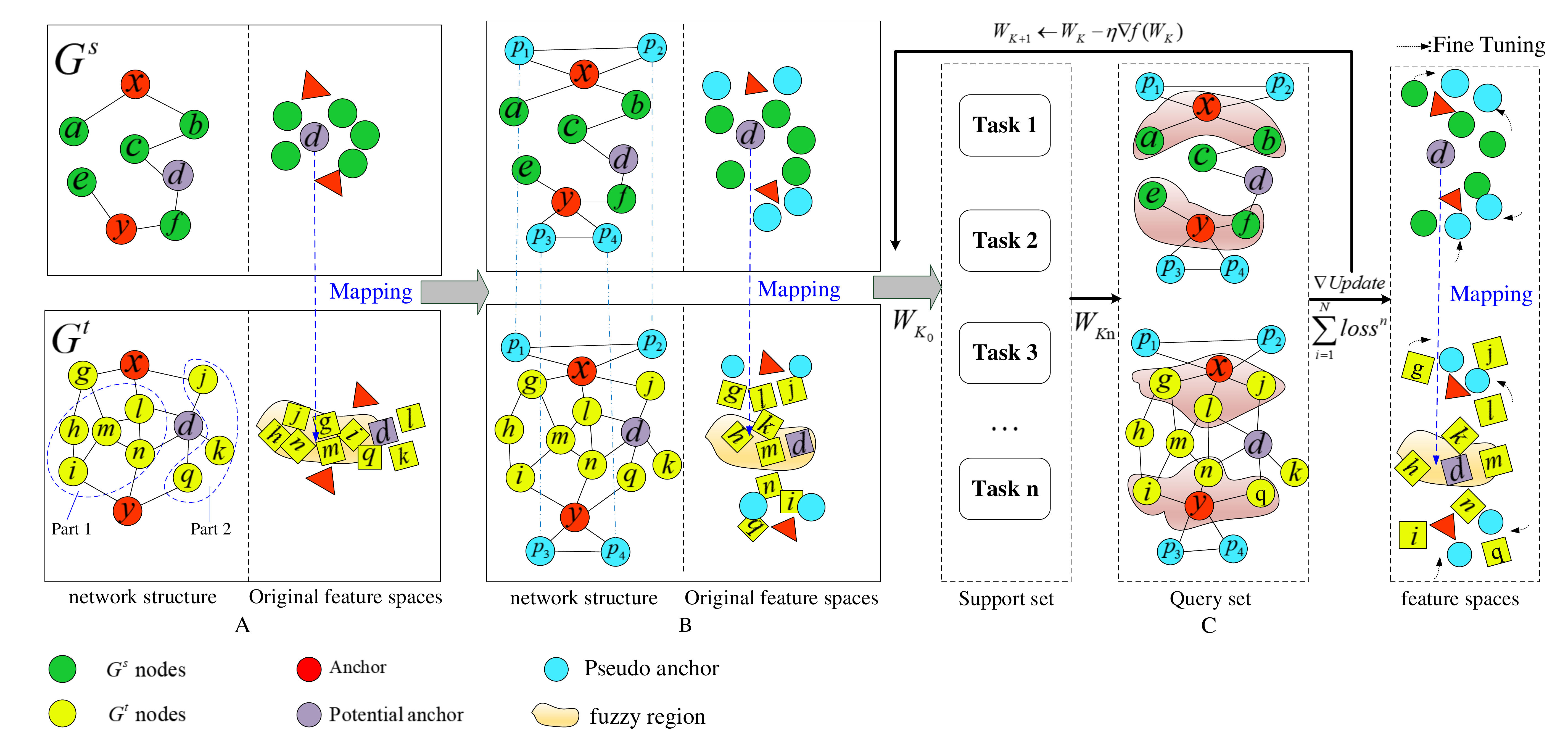} 
\caption{A toy example of the proposed framework. A. Structural proximity preserving objective sometimes could lead to ``overly-close'' embeddings for nodes in dense neighborhood structure during the embedding learning. B. Pseudo anchors help the corresponding embeddings learned to be more widely apart. C. The meta-learning step fine-tunes the positions of the pseudo anchors.}
\label{framework}
\end{figure*}

Fig. 1 illustrates the ``overly-close'' embedding issue. Consider the case shown in Part A of Fig. 1. There are two networks $G^s$ and $G^t$, in which $x$ and $y$ represent two observed anchor pairs. There is also a potential anchor $d$ to be aligned. 
Node $d$ has several first-order neighbors, and shares multiple common first-order neighbors ($l$, $n$, $q$, $j$) with its second-order neighbors ($m$, $x$ and $y$) in $G^t$, which means there is a relatively ``dense'' local structure centering $d$ within the second-order neighborhood. That is to say, $d$ itself, its first-order neighbors (i.e., $l$, $n$, $q$, $k$, and $j$) and its second-order neighbors (i.e., $m$, $x$, and $y$), form a dense clique with its average degree larger than that of the entire network.
Conventional structural proximity-based embedding algorithms \cite{DBLP:conf/kdd/PerozziAS14,DBLP:conf/www/TangQWZYM15,DBLP:conf/kdd/GroverL16} map a potential counterpart of $d$ in $G^s$ into a low-dimensional subspace. However, it is hard to distinguish $d$ from its neighbors within the clique due to the fact that the numerical representations of the nodes in a dense local structure are rather similar to each other in the embedding space, especially when there are not enough anchors to provide additional information for the alignment. In this paper, we will refer to the regions covering the embeddings of such cliques as ``fuzzy regions".

An intuitive idea to address the aforementioned issue is to restrict the number of nodes in the ``fuzzy regions'' by forcing the node embeddings to be more widely apart in the embedding space, and thus more distinguishable from each other. To achieve that,
we propose a two-fold strategy. 
First, we implant pseudo anchors by directly connecting to some real anchor(s) (See Fig. \ref{framework} Part B). Under the structural preserving objective, we expect that the pseudo anchors can have more influence to the local structure formed by the real anchors and their first-order neighbors, but contribute less impact on the nodes topologically far from the anchors. The enlarged group of the anchor's first-order neighbor will form a more compact clique thus to better distinguish from other nodes which is topological far from the anchor. By pulling this enlarged group of nodes away from other nodes in the learning process via pseudo anchors, the node embeddings learned can be enforced to be more widely apart in the embedding space.
Second, we try to address the challenge of determining the proper positions of pseudo nodes in the embedding space. Although implanting pseudo anchors enables users to have embeddings more widely apart, the negative influence will be introduced when an inappropriate embedding of the pseudo anchor is set in the ``fuzzy regions''. To this end, a fine-tuning strategy based on meta-learning is proposed in the learning process. The motivation is to learn a updating direction that can drive the embeddings of the pseudo anchors far from the ``fuzzy regions" in the learning process. In particular, we make the pseudo anchors across networks close and the real anchors' first-order neighbors farther from them. Since the anchors' first-order neighbors are around ``fuzzy regions'', pulling pseudo anchors far away from them can ensure the pseudo anchors to be updated in the right direction. This strategy at the same time can alleviate the ``overly-close'' issue of the enlarged group around the anchors that could be introduced by implanting pseudo anchors (Refer to Subsection 3.2 for details). As shown in Fig. \ref{framework} Part C, under the supervision of the observed anchors and the prior knowledge (learned from support datasets with rich labeled data), pseudo nodes' positions are adjusted during the model learning process. Finally, a more discriminative embedding space can be obtained for user alignment across social networks.

The contributions of this paper are summarized as follows:
\begin{itemize}
\item We identified that overly-close embeddings among anchors and their neighbors, which is encountered in most embedding based alignment models, is one of the key issues causing the failure alignment results.
\item We propose implanting pseudo anchors as a mechanism to allow user embeddings learned to be more widely apart in the embedding space.

\item We propose a unified learning framework which with the use of pseudo anchors and a meta-learning scheme can be applicable to the existing embedding-based alignment models for learning better organized embedding spaces to achieve high alignment accuracy.

\item We evaluate the proposed framework by applying it to several state-of-the-art embedding based models. The experimental results demonstrate that the framework is effective in boosting the accuracy of embedding based alignment models 
compared with their original versions.
\end{itemize}

\section{Related Work}
\subsection{Network embedding}

Network embedding aims at mapping each node of a network into a low dimension space with the structural proximity property preserved. Earlier works in this area are mainly based on dimension reduction via matrix  factorization\cite{DBLP:journals/tnn/HouZXZL09, DBLP:conf/nips/BelkinN01,DBLP:conf/kdd/YanHJ09}.
Some more recent ones like GraRep \cite{DBLP:conf/cikm/CaoLX15} and HOPE \cite{DBLP:conf/kdd/OuCPZ016} try to decompose multiple types of matrices defined for modeling high-order and global structural information.  

Inspired by the Skip-gram algorithm of word2vec \cite{DBLP:journals/corr/abs-1301-3781}, shallow neural network based algorithms, such as DeepWalk \cite{DBLP:conf/kdd/PerozziAS14} LINE  \cite{DBLP:conf/www/TangQWZYM15} and Node2vec \cite{DBLP:conf/kdd/GroverL16}, were proposed for network embedding. Negative sampling and Stochastic Gradient Descent (SGD) adopted in the optimization make the embedding learning scalable. The main difference among them resides on how they determine the  ``context" for the target nodes. Depth First Search (DFS) \cite{DBLP:conf/kdd/PerozziAS14}, Breadth First Search (BFS) \cite{DBLP:conf/www/TangQWZYM15}, and the balance strategy between DFS and BFS \cite{DBLP:conf/kdd/GroverL16} are leveraged in them respectively. By drawing the equivalence between Skip-gram and matrix factorization, NetMF \cite{DBLP:conf/wsdm/QiuDMLWT18} was proposed for unifying network representation learning into a matrix factorization framework. Further, NetSMF \cite{DBLP:conf/ijcai/ZhangDWTD19} and ProNE \cite{DBLP:conf/www/QiuDMLWWT19} address the efficiency issues in the matrix factorization framework by leveraging sparse matrices for fast and scalable learning.

To achieve powerful representation capability, deep neural networks have also been proposed for network embedding.
For instance, SDNE \cite{DBLP:conf/kdd/WangC016} uses a deep autoencoder to maintain the similarity between the first and second orders. GraphGAN \cite{DBLP:conf/aaai/WangWWZZZXG18} introduces the GAN model with the use of a generator to fit the real distribution of the node embedding. 
Also, Convolutional Graph Neural Networks (ConvGNNs) have attracted extensive attention for modeling graphs. For instance, Graph Convolutional Network (GCN) \cite{DBLP:conf/iclr/KipfW17} uses the eigenvalues and eigenvectors of the Laplacian matrix to learn the graph properties. Furthermore, several works, such as GAT \cite{DBLP:journals/corr/abs-1710-10903} and RWNN \cite{DBLP:conf/iclr/RongHXH20}, incorporate the attention mechanism into the GNN model for more robust learning.
In addition, to address the learning issues related to ConvGNNs, GraphSAGE \cite{DBLP:conf/nips/HamiltonYL17} aggregates randomly selected neighboring nodes for efficient learning, and DROPEDGE \cite{DBLP:conf/iclr/KimDOA21} learns the representation by randomly dropping edges for alleviating the over-smoothing problem. HNN \cite{DBLP:conf/www/SunYLCMHC21} adopts hyperedge distillation for hpyergraph based representation learning. For all these models, even with a deeper architecture utilized, the strategy for aggregating neighbors' information in each layer is essentially a structural preserving process. 

For structural preserving embedding algorithms applied to a single social network, the objective function tends to make neighboring nodes (such as adjacent nodes or nodes that share many common neighbors) as close as possible. This is reasonable for some downstream tasks such as link prediction, as there is a high probability for users being friends if they are in ``structural'' proximity. However, for user alignment across social networks, we expect to  learn a space not only can maintain the structural proximity but also allow the node embeddings to be organized more evenly apart to ease the identification of the corresponding anchor from its close neighbors in another network. Therefore, how to learn the network embedding driven by the objective of the alignment task remains open, which is the main goal of this paper.

\subsection{Network Alignment} 
Recent studies on social network alignment can roughly be categorized as unsupervised and (semi-)supervised. Unsupervised social network alignment assumes absence of anchor labels and considers the task as a general graph alignment problem. Some representative methods include BIG-ALIGN \cite{DBLP:conf/icdm/KoutraTL13} which performs the alignment based on Alternating Projected Gradient Descent, and UMA \cite{DBLP:conf/icdm/ZhangY15} which suggested the use of a two-step matrix factorization. Recently, embedding based algorithms have been widely adopted for the unsupervised alignment. They learn user representations from the user social relationships and/or the user profiles. 
Co-training \cite{zhong2018colink} and iterative learning \cite{zhou2018FRUIP} can be adopted for boosting alignment performance. 
Besides, REGAL \cite{heimann2018regal} was proposed with a similarity matrix factorization introduced for effective representation learning, which is also applicable to multiple networks. Factoid Embedding (FE) \cite{DBLP:conf/icdm/XieMLZL18} models user relationships and profiles as a knowledge graph, and learns a unified embedding space for the alignment. CONE-Align \cite{DBLP:conf/cikm/ChenHVK20} uses a multi-granularity strategy for the alignment via mapping of the graph structure and users. Instead of aligning the users one by one, Li et al. \cite{DBLP:conf/cikm/LiWYZZLL18} consider all the users in a social network as a whole and perform user alignment from the user distribution level. They proposed UUIL$_{gan}$ and UUIL$_{omt}$ models based on earth mover’s distance. Experimental results show  comparative performance with supervised baselines.

Due to the high structural complexity and the fact that the presence of user profiles cannot be always assumed in different social networks, generalization of these unsupervised models are unstable. So far, the use of semi-/supervised methods for the alignment is still the mainstream.

Semi-/supervised alignment methods assume that some anchor labels which refer to some aligned users are known to facilitate the alignment of the others. Traditional classification based methods \cite{DBLP:conf/cikm/KongZY13,
DBLP:conf/kdd/ZafaraniL13} was first proposed to predict if two users in different networks should be aligned.
Embedding based methods were later on proposed with superior performance and efficiency.
They mainly follow two approaches: embedding sharing and embedding mapping.

Embedding sharing methods learn a unified embedding space to achieve the alignment goal via sharing the embeddings of labeled anchors in different networks. For instance, IONE \cite{DBLP:conf/ijcai/LiuCLL16} learns the embedding space by preserving second-order follower-ship/followee-ship proximity with the embeddings of the anchors shared. Besides, structural diversity was further considered in its extension IONE-D \cite{DBLP:journals/tkde/LiuLCL20}. 
ABNE \cite{DBLP:journals/access/LiuZFZHZ19} utilizes an attention mechanism to obtain more robust alignment. CrossMNA \cite{DBLP:conf/www/ChuFYZHB19} leverages the cross-network information to refine ``inter-vectors'' and ``intra-vectors'' for aligning users across multiple networks. DALAUP \cite{DBLP:conf/ijcai/ChengZY0LT019} adopts active learning where the user embeddings are updated iteratively by active sampling and anchor user prediction. Motivated by the inherent connection between hyperbolic geometry and social networks \cite{DBLP:conf/kdd/ZhangCMCYX21}, the studies \cite{DBLP:conf/aaai/SunZZWPSY21,wang2020hyperbolic,sun2020perfect} proposes to study a very interesting problem, i.e., user identity linkage based on hyperbolic geometry. For instance, the distance between nodes can be defined using the Lorentz model \cite{wang2020hyperbolic} and Poincar{é} ball model \cite{sun2020perfect}. Then, the random walk-based structural proximity objectives can be optimized for the alignment. 

Rather than learning a unified embedding space for users across networks, embedding mapping methods learn also a mapping between separated embeddings for the alignment. For instance, based on the spaces learned by preserving the first-order proximity in individual networks, PALE \cite{DBLP:conf/ijcai/ManSLJC16} employs MLP (Multi-Layer Perceptron) to conduct a supervised latent space matching. Deeplink \cite{DBLP:conf/infocom/0002LZTWZ18} uses a deep neural network-based dual learning process to achieve a more powerful alignment model. SNNA \cite{DBLP:conf/aaai/LiWWYLLL19} and MSUIL \cite{DBLP:conf/cikm/LiWWLYLW19} learn the projection function which minimizes the Wasserstein distance between the anchors' distributions of two social networks. dName \cite{DBLP:conf/infocom/0002WTZZL19} adopts a disentangled Graph Convolutional Network embedding algorithm to iteratively aggregate feature information from local graph neighborhood. MGCN \cite{DBLP:conf/kdd/ChenYS0GM20}, considering the simple network topology information insufficient, conducts convolution on both local network structure and hypergraph network structure, and then minimizes the distance between anchors to learn the embedding mapping. For studying the robustness of alignment model, the studies \cite{DBLP:conf/ijcai/SuSZLQ18,DBLP:journals/tkde/ZhangSSQL21} attempt to reconcile multiple social networks by comprehensively exploiting attribute and structure information via an embedding approach. By observing that social network alignment and behavior analysis can benfit from each other, BANANA \cite{DBLP:conf/ijcai/RenZZSSZG20} takes the first attempt to study the joint problem of social network alignment and user behavior analysis.

Different from the aforementioned works, our approach focuses on learning the user embedding space so that the users are more evenly distributed
via explicit introduction of pseudo anchors. This approach has the advantage that we need only slight changes to the original network structure and use meta-learning to guide the updating direction of implanted pseudo anchors. Since we do not need to modify the learning algorithms of the original model, the proposed framework can be applied on top of different embedding based alignment models to further boost their alignment performance.

\section{Model Framework}

Given two different social networks as the source network $G^s=(v^s,e^s)$ and the target network $G^t=(v^t,e^t)$ respectively where $v^s$ and $v^t$ denote the users in $G^s$ and $G^t$, and $e^s$ and $e^t$ denote the edges within each network. We define a function $\Phi_{v_{i}}: v_{i}\rightarrow \overrightarrow{u}_{i}$ which embeds node $v_i$ into a low-dimensional vector $\overrightarrow{u}_{i}$ under the structural proximity assumption. The network alignment problem is here defined as a mapping function $f_m(\overrightarrow{u}^s_i,\overrightarrow{u}^t_j)\in \{0,1\}$ and at the same time the node embeddings based on the supervision of some given anchor nodes. 

With the objective to design methodologies which are generally applicable to a wide spectrum of embedding-based network alignment algorithms, 
we propose a learning framework which can increase the alignment accuracy by ensuring the inferred embeddings to possess desirable properties for the alignment. In particular, we propose to ``implant'' \textit{pseudo anchors} as the means for steering the learning of the pseudo anchors' embeddings to avoid the embeddings of nodes in the neighborhood of anchors to be closely clustered (named as \textit{fuzzy regions} in Section 1), and thus affecting the alignment accuracy. We achieve the goal via a two-fold strategy. The first is to pull nodes directly connected with the real anchors farther away from the fuzzy regions via the intervention of pseudo anchors. The second is to use meta-learning to adjust the updating directions of the pseudo anchors. This can ensure that the implanted pseudo anchors will not approach to the fuzzy regions during learning and can again avoid the ``overly-close'' embeddings to be formed around real anchors due to the introduction of the implanted pseudo anchors. The proposed framework can learn an embedding space in which the node embeddings are more evenly distributed, and can be seamlessly integrated with different network alignment methods for learning a better mapping across networks. 

\subsection{Implanting Pseudo Anchors in Networks}

\begin{figure*}[ht]
\centering
\includegraphics[width=0.9\linewidth]{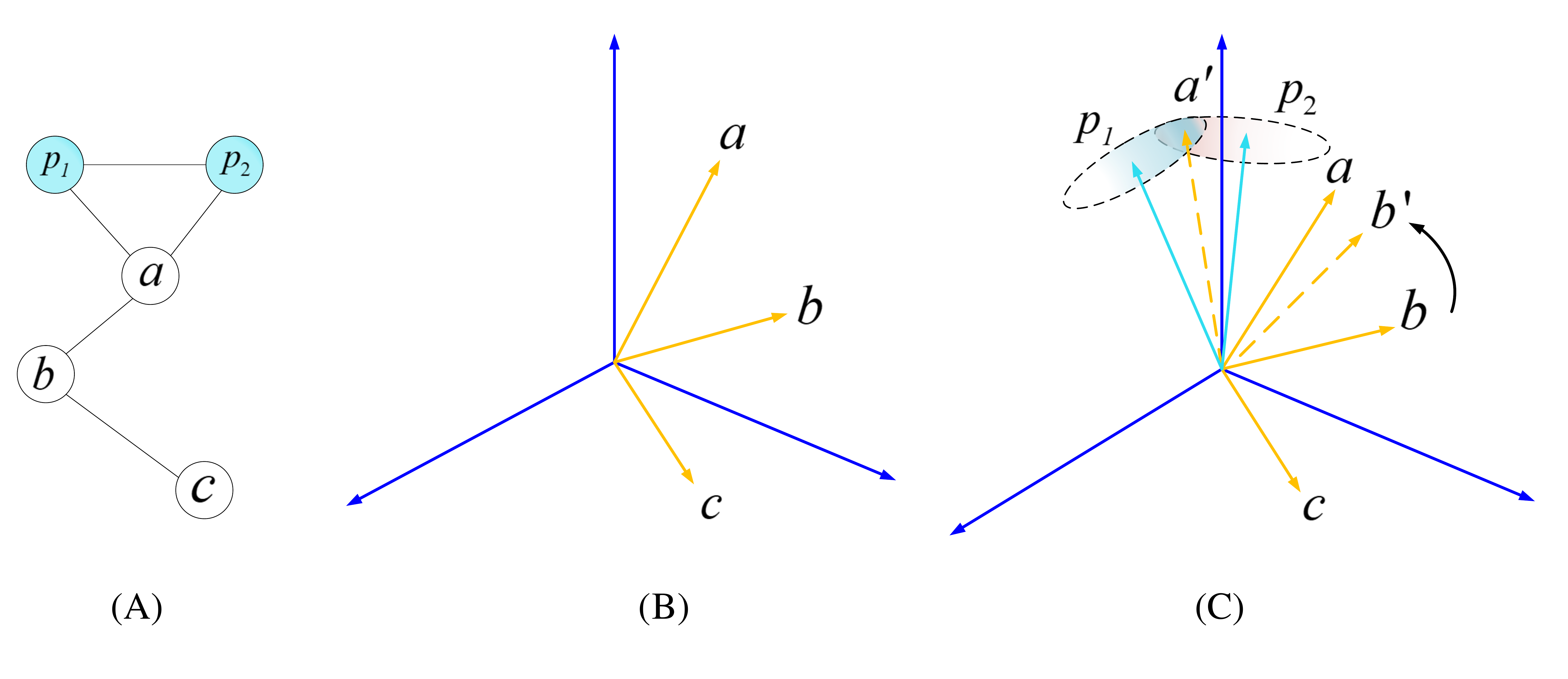}  
\caption{Illustration of embedding shifting when pseudo anchors are implanted. A. Network structure after implanting pseudo anchors. B. Embedding distribution before implanting pseudo anchors. C. Embedding distribution distribution after implanting pseudo anchors.}
\label{PsedoAnchorSparse}
\end{figure*}

Most embedding-based network alignment methods considers the structural preserving objective for learning the node embeddings. To learn better node embeddings with minimial modifications to the objective function, we propose to implant pseudo anchors to result in a more even distribution of nodes in the embedding space.
The key idea is to implant the pseudo anchors to exercise more influence to the local structures around real anchors, with less impact on nodes far from the anchors in the embedding space. Then, nodes in the neighborhood of anchors can have their inferred embeddings farther apart from each other under the pulling effect of the pseudo anchors (See Fig. \ref{framework} Part B). Therefore, the ``overly-close" phenomenon around anchor nodes can be alleviated.

To further explain why implanting pseudo anchors can result in an even distribution of embeddings for nodes in the higher-order neighborhood of the real anchors, 
we take the learning process of typical structural preserving embedding algorithms \cite{DBLP:conf/kdd/PerozziAS14,DBLP:conf/www/TangQWZYM15,DBLP:conf/kdd/GroverL16}\footnote{The difference between these algorithms is the way of determining ``context" nodes for the target node.} as an example. 
The objective of these algorithms tends to embed a certain node $v_i$ and its adjacent nodes in the local structure as close as possible in the embedding space, and keep $v_i$ far away from random sampled nodes simultaneously. Here we refer $v_i$'s adjacent nodes and random sampled nodes as $context(v_i)$ and $neg(v_i)$ separately. Then for node $v_i$ and $v_j \in  \left\{ context(v_i)  \cup  neg(v_i) \right\}$, the structural preserving objective between them can be written as:
\begin{equation}
\begin{aligned}
\label{L_emb}
    \mathcal{L}(v_i,v_j)\!=\!L^i_{j}log[\sigma({\overrightarrow{u}_i}^T \overrightarrow{u}_{j})]\!+\!(1\!-\!L^i_{j})\log[1\!-\!\sigma({\overrightarrow{u}_i}^T \overrightarrow{u}_{j})]
\end{aligned}
\end{equation}
where $\overrightarrow{u}_i$ and $\overrightarrow{u}_{j}$ are the corresponding embeddings for $v_i$ and $v_j$
and $\sigma(.)$ is the sigmoid function. The value of $L^i_j$ depends on whether $v_j$ is the ``context node'' of $v_i$.
\begin{equation}
L^i_j=\begin{cases} 
1,& v_j \in context(v_i) \\
0,& v_j \in neg(v_i)
\end{cases}.
\end{equation}
We can then compute the gradient of $\overrightarrow {u}_i$ for model optimization, given as:
\begin{equation}\label{UP_ui}
\frac{\partial{\mathcal{L}(v_i,v_j)}}{\partial{\overrightarrow {u}_i}}=[L^i_j-\sigma({\overrightarrow {u}_i}^T \overrightarrow {u_{j}})] \overrightarrow {u_{j}}
\end{equation}
For a specific anchor node $v_a$, the updating rule of $\overrightarrow{u}_a$ becomes:
\begin{equation} \label{ori_updating_anchor}
\begin{aligned}
    \overrightarrow{u}_{a}:=\overrightarrow {u}_{a}+
    & \eta
    \sum_{v_{j} \in \left\{ context(v_a) \cup neg(v_a) \right\}}\frac{\partial{\mathcal{L}(v_a,v_{j})}}{\partial{\overrightarrow{u}_a}}
\end{aligned}
\end{equation}
Note that $v_a$ and $v_i$ are exchangeable as here $v_a$ is the target node to be updated. We can replace $v_i$ with $v_a$ in Eq. (\ref{UP_ui}) for computing the corresponding updating.

Then, consider the case of implanting pseudo anchors shown in Fig. \ref{PsedoAnchorSparse}.A where $v_a$ is the anchor node and $\{ p_{1}, p_{2} \}$ are the implanted pseudo anchors. The anchor $v_a$ now has to satisfy not only the first-order approximation with node $b$, but also that with the pseudo anchors. Therefore, $\{p_1,p_2\}$ should be included in context of $v_a$, and the updating rule of $\overrightarrow{u}_{a}$ becomes:
\begin{equation}\label{Update_ua}
\begin{aligned}
\overrightarrow{u}_{a}:=\overrightarrow{u}_{a}+& \eta  
     {\sum_{v_j \in \left\{ neg(v_a)  \cup  \atop context(v_a)\cup \{p_1, p_2 \} \right\}}\frac{\partial{\mathcal{L}(v_a,v_j)}}{\partial{\overrightarrow{u}_{a}}}} 
\end{aligned}
\end{equation}

Compared with the case without implanting pseudo anchors,  $v_a$'s embedding will be shifted from $a$ to $a'$ as shown in Fig. \ref{PsedoAnchorSparse}.C where the shifting  $\Delta{\overrightarrow{u}_a}$ is derived by subtracting Eq. (\ref{Update_ua}) from Eq. (\ref{ori_updating_anchor}), given as:
\begin{equation}\label{shifta}
    \Delta{\overrightarrow{u}_a}= \eta (\frac{\partial{\mathcal{L}(v_a,v_{p_1})}}{\partial{\overrightarrow{u}_a}}+
    \frac{\partial{\mathcal{L}(v_a,v_{p_2})}}{\partial{\overrightarrow{u}_a}})
\end{equation}
According to Eq.(\ref{shifta}), in order to maintain the first-order proximity between anchors $a$ and $p_{1}$, $p_{2}$,
$\overrightarrow{u}_{a}$ will approach towards  $\overrightarrow{u}_{p_{1}}$ and $\overrightarrow{u}_{p_{2}}$.
The shifting of anchor $v_a$’s embedding will also affect anchor $v_a$’s first-order neighbor node $v_b$. Based on the structure illustrated in Fig. \ref{PsedoAnchorSparse}.A and the updating rule of Eq. (\ref{UP_ui}), 
the shifting of $v_b$ is given as:
\begin{equation} \label{shiftb}
    \Delta{\overrightarrow{u}_b} \propto \sigma({\overrightarrow {u}_b}^T (\overrightarrow {u_{a}}+\Delta{\overrightarrow{u}_a}))-\sigma({\overrightarrow {u}_b}^T \overrightarrow {u_{a}})
\end{equation}
Similarly, the shifting of $v_c$ can be derived as:
\begin{equation} \label{shiftc}
    \Delta{\overrightarrow{u}_c} \propto \sigma({\overrightarrow {u}_c}^T (\overrightarrow {u_{b}}+\Delta{\overrightarrow{u}_b}))-\sigma({\overrightarrow {u}_c}^T \overrightarrow {u_{b}})
\end{equation}

In general, the pulling effect caused by the pseudo anchors will propagate out via the anchors' higher-order neighborhood.
As the order of proximity increases, the amount of shifting will decrease accordingly\footnote{We provide the proof in Appendix A}. Therefore, nodes that are farther away from the anchors will be shifted less. Based on this, we can conclude that pseudo anchors have a high impact on the corresponding anchor nodes and neighbors that close to them. 
Since the nodes around anchors are pulled away by pseudo anchors, the embeddings of higher-order neighbors will be more evenly distributed and thus easier to be distinguished. 

With the effect of implanting pseudo anchors and the potential benefits explained, 
the remaining key issue is how to properly initially place the pseudo anchors and then update them during the learning process. 
It is not too difficulty to notice that improperly updating the pseudo anchors (e.g., moving them towards the ``fuzzy regions'') will lead to undesirable embedding results (e.g., all nodes embedded in one ``over-closely'' cluster), and in turn causing negative effects to the alignment task. To this end, we try to determine the proper position of implanted pseudo anchors in the learning process. Specifically, we introduce a meta-learning approach to control the updating direction of pseudo anchors as described in the next subsection.

\subsection{Fine-Tuning of Pseudo Anchors }

To ensure implanting pseudo anchors to be properly controlled to realize their benefits as explained, we propose a fine-tuning meta-learning strategy 
with two goals to be achieved. The first is to ensure the updating directions of pseudo anchors to be far away from the ``fuzzy regions''. The second is to avoid ``overly-close" embeddings around the anchors caused by the implanted pseudo anchors. To achieve these goals, we first utilize a meta-learning based algorithm to learn some prior knowledge about the updating from some support datasets that contain rich labeled anchors. 
We then make use of the prior knowledge for fine-tuning the pseudo anchors in each embedding learning epoch. Through the interleaved iterations of the fine-tuning and embedding learning steps, we expect to update the pseudo anchors in a properly controlled manner, and thus to result in a more evenly distributed embedding space. In the following, we provide the details of the proposed algorithm.

Given a specific anchor user $v_{a}$, let $P_a=\{p^0_{a},p^1_{a},...,p^n_{a}\}$ be the set of implanted pseudo anchors corresponding to $v_{a}$, where the superscript denotes the index of the pseudo anchor. For a specific pseudo anchors $p^i_{a}$, we define the its updating direction as:
\begin{equation}
    \Delta{\overrightarrow{p}_{a}^i}(v_{a},W)=g\left( w^i\frac{\overrightarrow{u}_{a}+{\sum_{j\in{nei(v_{a})}}\overrightarrow{u}_j}}{N+1}\right).
    \label{DirectionControl}
\end{equation}
where $W=\{ w^0, w^1,...,w^n \}$ is the set of learnable parameters used for controlling the updating directions of the pseudo anchors, 
$nei(v_a)$ denote the first-order neighbors of $v_a$, and $g$ is the activation function.

\begin{figure}[htbp]
\centering
\includegraphics[width=1\linewidth]{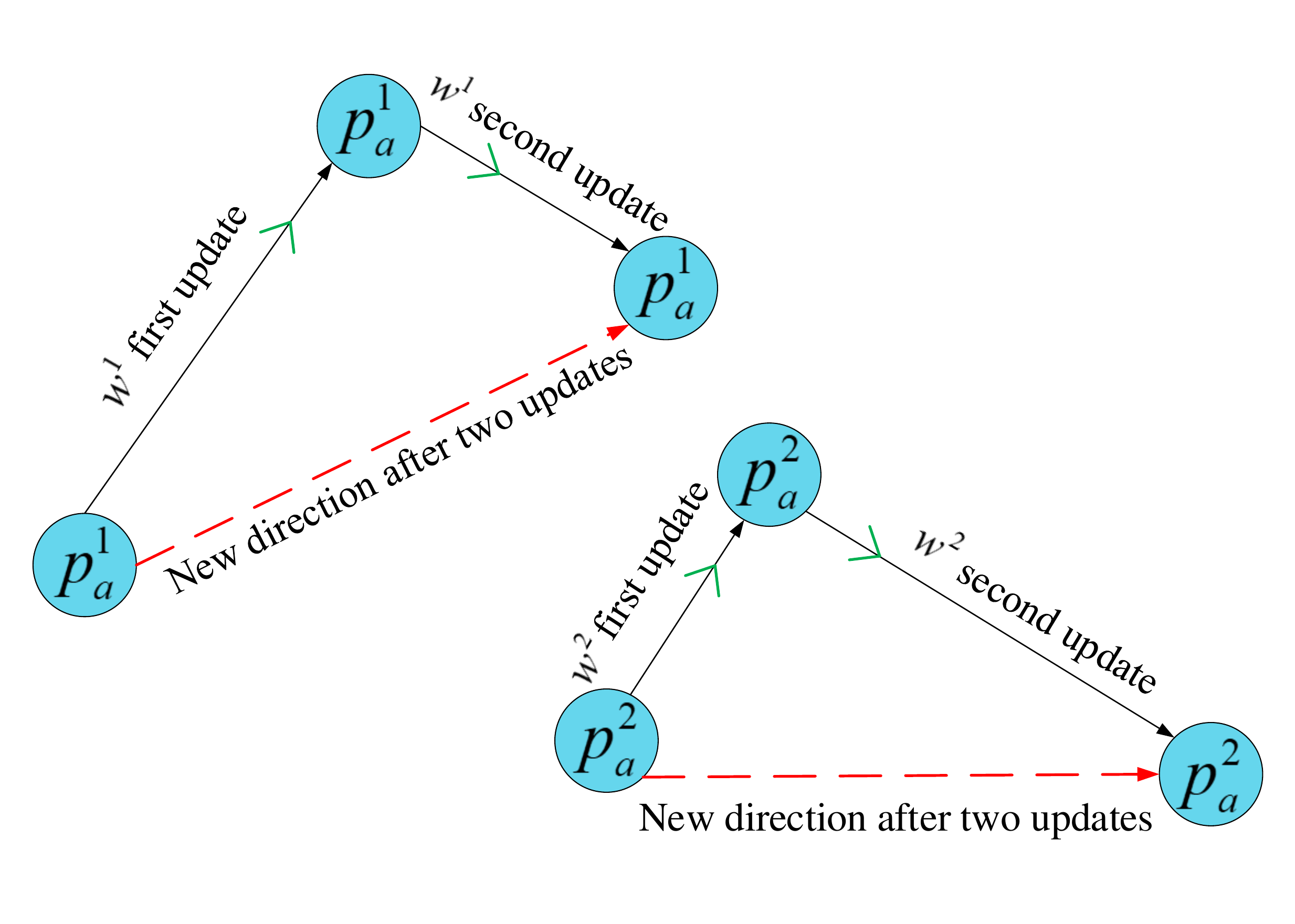}
\caption{Illustration of Controlled Direction Updating}
\label{two_iter}
\end{figure}

According to Eq.(\ref{DirectionControl}), the pseudo anchors connecting the same real anchor will have the same ``base direction'' $\overrightarrow{u}_a+\sum_{j\in{nei(v_{a})}}\overrightarrow{u}_j$ which control the updating directions. By setting different values to the coefficient $w^is$ of ``base direction" and applying it iteratively, different directions of updating pseudo anchors can be resulted. Take Fig. \ref{two_iter} as an example. $p_a^1$ and $p_a^2$ are pseudo anchors implanted to the same anchor $v_a$ and the updating direction is determined by applying the ``base direction'' twice with the corresponding parameters $w^1$ and $w^2$. We can see that, although the directions for the two iterations of updating $p_a^1$ and $p_a^2$ are the same, the different updating step length controlled by $w^1$ and $w^2$ can lead to different updating directions for different pseudo anchors.

\begin{figure}[ht]
\centering
\includegraphics[width=1\linewidth]{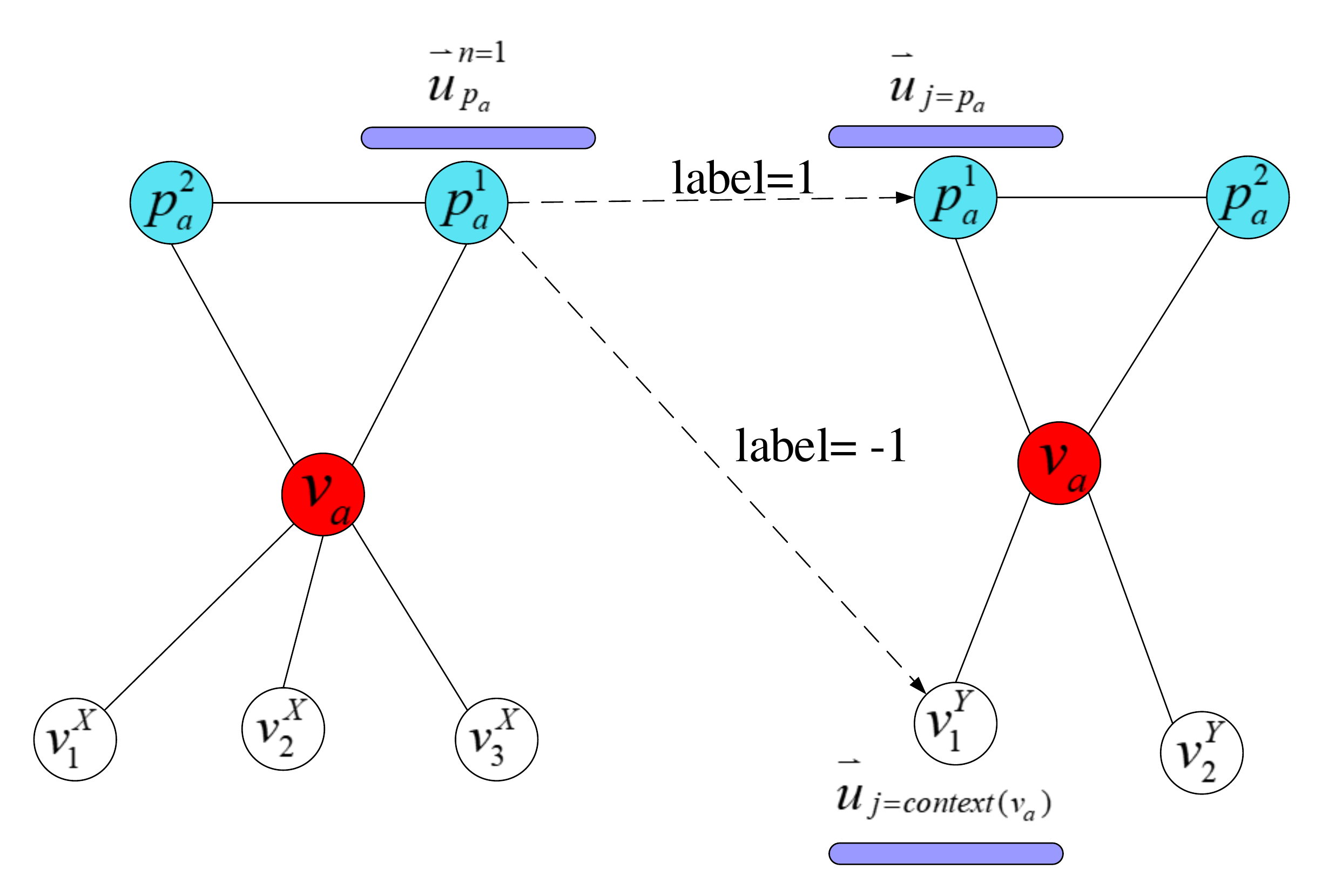}
\caption{Basic idea of fine-tuning}
\label{FineTuning}
\end{figure}

Learning such $W$ is challenging when the labeled anchors are insufficient which are generally true in many cases. 
So we try to leverage the data from other networks in which anchors can easily be obtained as the support dataset $S={\lbrace S_{1}, S_{2}, S_{3}...S_{K} \rbrace}$ for learning $W$. We consider the to-be-aligned networks as the query set $Q$. We first learn  $W$ from the support set as prior knowledge. We then calculate the total loss based on the $W$ transferred from the support set and updating $W$ according to the total loss in the query set.
In particular, for the first step, we try to coincide the position of pseudo anchors across networks and at the same time make them far away from first-order neighbors of real anchors. We define the objective function as:
\begin{eqnarray}
    f(U_{S_i})&=-\sum\limits_{p_a^n\in{P_a}}\sum\limits_{v_j \in \{ \{p_a\}\cup  {context\{v_a\}\}}}^{m} \nonumber\\
    &( \frac{1}{1+e^{-label\cdot (\overrightarrow{u}_{p_a}^n+ \Delta {\overrightarrow{p}_{a}^n}(v_{a},W))\cdot{\overrightarrow{u}_{j}}}}) \nonumber\\
    & where \ \overrightarrow{u}_{p_a}^n,\overrightarrow{u}_{j} \in U_{S_i}
    \label{ObjectiveFunction}
\end{eqnarray}
where $\overrightarrow{u}_{p_a}^n$ is the embedding of $n-{th}$ pseudo anchor in $P_a$, $U_{S_i}$ is the embedding space of the support data $S_i$. $\overrightarrow{u}_{j}$ can correspond to pseudo anchor $p_a$ or the first-order neighbor of the real anchor associated with $p_a$ in the other network. The value of $label$ is $1$ when  $\overrightarrow{u}_{p_a}^n$ and $\overrightarrow{u}_{j}$ correspond to an anchor pair, or $-1$ otherwise. This objective function makes pseudo anchor pairs across networks close in the embedding space and push the pseudo anchors away from the direct neighbors of the real anchors. The intuition of this idea is illustrated in Fig. \ref{FineTuning}. Rather than implanting pseudo anchors making nodes in the ``fuzzy regions'' to be more evenly distributed, Eq. (\ref{ObjectiveFunction}) aims at alleviating the ``overly-close'' group problem around the anchors. It guarantees implanting pseudo anchors will not cause negative effect to the alignment. 
Then, based on the support set $S_{i}$ in $S={\lbrace S_{1}, S_{2}, S_{3}...S_{K} \rbrace}$, the corresponding parameter $W$ can be updated as:
\begin{equation}
    W=W-\eta_1 \bigtriangledown f(U_{S_i})
    \label{update_support}
\end{equation}
where $U_{S_i}$ is the embedding space of $S_i$, $\eta$ is the learning rate.

We then transfer the learned $W$ to query set $Q$ (dataset with less anchor labels). We define the objective function based on the transferred $W$ (learned from the support set) as:
\begin{eqnarray}
    f(U_{Q};W)&=-\sum\limits_{p_a^n\in{P_a}}\sum\limits_{v_j \in \{ \{p_a\}\cup  {context\{v_a\}\}}}^{m} \nonumber\\
    &( \frac{1}{1+e^{-label\cdot (\overrightarrow{u}_{p_a}^n+ \Delta {\overrightarrow{p}_{a}^n}(v_{a},W))\cdot{\overrightarrow{u}_{j}}}})
    \label{ObjQuery}
\end{eqnarray}
where $\overrightarrow{u}_{p_a}^n, \overrightarrow{u}_{j} \in U_{Q}$. For the first iteration, we use transferred $W$ to initialize the $\Delta {\overrightarrow{p}_{a}^n}(v_{a},W)$ in $f(U_{Q};W)$. $W$ can then be updated as:
\begin{equation}
    W=W-\eta_{2} \bigtriangledown f(U_{Q};W).
    \label{updating_rule}
\end{equation}
We repeat this process $K$ times by using $W$ in last step to calculate $\Delta {\overrightarrow{p}_{a}^n}(v_{a},W)$ iteratively. Finally, we obtain the direction for the updating of the pseudo anchors. Algorithm 1 summarizes the key steps of fine-tuning the pseudo anchors.

\begin{algorithm}[htbp]
\caption{Meta-Learning for Fine-Tuning the Pseudo Anchors}
\algorithmicrequire \ Structural preserving embedding algorithm $\Phi$, the initial weights $W$, support set $S$, query set $Q$,  the learning rate $\eta_{1}$, $\eta_{2}$, iteration number $K$.\\
\algorithmicensure \ The learned parameters $W=\{ w^0, w^1,...,w^n \}$  for fine tuning.
\begin{algorithmic}[1]
\For{$S_i \in  S$}
\State Obtain the embedding space $U_{S_i} \sim  \Phi(S_i)$
\State Compute $\Delta{\overrightarrow{p}_{a}^i}(v_{a},W)$ according to Eq. (\ref{DirectionControl})
\State $W \leftarrow W-\eta_{1}\bigtriangledown f(U_{S_i}) $,  $f(U_{S_i})$ can be obtained according to Eq. (\ref{ObjectiveFunction}) 
\EndFor

\State Obtain the embedding space $U_{Q} \sim  \Phi(Q)$

\For{$k \in [1,K]$}
\State Compute $\Delta{\overrightarrow{p}_{a}^i}(v_{a},W)$ according to Eq. (\ref{DirectionControl})
\State $W \leftarrow W-\eta_{2}\bigtriangledown f(U_{Q};W) $,  $f(U_{Q};W)$ can be obtained according to Eq. (\ref{ObjQuery}) 
\EndFor
\State Obtain the final direction of pseudo anchors $\Delta{\overrightarrow{p}_{a}^i}(v_{a},W)$ according to Eq. (\ref{DirectionControl}) 
\State Update pseudo anchor according to $\overrightarrow{u}_{p_a}^n+ \Delta {\overrightarrow{p}_{a}^n}(v_{a},W)$
\end{algorithmic}
\end{algorithm}

\section{EXPERIMENT AND ANALYSIS}
\subsection{Dataset Description}
To evaluate the performance of the proposed framework, we conduct experiments on two real-world datasets. The first one is Twitter-Foursquare \cite{DBLP:conf/ijcai/ZhangY15,DBLP:conf/ijcai/LiuCLL16,DBLP:journals/access/LiuZFZHZ19,DBLP:journals/tkde/LiuLCL20}, a widely used dataset in the literature. Users are collected from two famous social networks, and the ground-truth anchors can be obtained as some Foursquare users provide their  Twitter accounts in their profiles. The second one is DBLP\footnote{https://www.aminer.cn/citation} \cite{DBLP:journals/tkde/LiuLCL20}. In this dataset, authors are split into different co-author networks by filtering publication venues of their papers. The first network contains authors who published papers in ``Data Mining'' related conferences or journals including SIGKDD, PAKDD, TKDD, etc. The other one consists of authors who published papers in ``Machine Learning'' related venues such as NIPS, ICML, ICONIP, etc. Ground-truth anchors are labeled as authors who published papers in both areas. Table \ref{dataset} lists the statistics of the two datasets.

Since the meta-learning algorithm we proposed requires a support set for learning the weights for updating pseudo anchors, we use an external dataset arXiv \cite{DBLP:conf/www/ChuFYZHB19} as it has rich labeled anchors compared with the Twitter-Foursquare and DBLP. It contains 13 sub-networks in terms of different arXiv categories, and the average ratio between every two sub-networks is 41.67\%.

\begin{table}[!htbp] \centering 
	\caption{Statistics of the datasets used for evaluation}
	\label{dataset}
	\begin{tabular}{c||c|c|c}
		\hline
		\hline
		Networks&\#Users&\#Relations&\#Anchors\\
		\hline
		\hline
		Twitter&5,220&164,919&\multirow{2}{*}{1,609}\\
		\cline{2-3}
		Foursquare&5,315&76,972&\\
		\hline
		DBLP\_DataMining&11,526&47,326&\multirow{2}{*}{1,295}\\
		\cline{2-3}
		DBLP\_MachineLearning&12,311&43,948&\\
		\hline
	\end{tabular}
\end{table}

\subsection{Baseline Methods}
Rather than designing a specific model, our proposed framework focuses on learning an ``alignment-oriented'' embedding via the intervention of pseudo anchors. It can be seamlessly incorporated into several embedding based alignment models. In this paper, we apply our framework to the following embedding based models:

\begin{itemize}
\item \textbf{IONE} \cite{DBLP:conf/ijcai/LiuCLL16} is a semi-supervised embedding alignment model in which follower and followee relationships are explicitly represented as input context and output context vectors. By preserving the second-order proximity, it learns user latent space under the supervision of partially labeled anchors.

\item \textbf{DEEPLINK} \cite{DBLP:conf/infocom/0002LZTWZ18} applies deep learning to social network alignment where random walk and Skip-gram algorithm are utilized for the embedding learning. Neural network based mapping and dual learning were proposed for the user alignment.

\item \textbf{ABNE} \cite{DBLP:journals/access/LiuZFZHZ19} is an attention-based network embedding model for user alignment. An attention mechanism is adopted for learning the alignment task-driven weights between users. User alignment is achieved by sharing the parameters between the network embedding based model and the attention mechanism.

\item \textbf{SNNA} \cite{DBLP:conf/aaai/LiWWYLLL19} is a weakly supervised model for user alignment. Based on users' embeddings learned using shallow neural networks, it introduces distribution closeness and adversarial learning to learn the mapping of users across networks.

\item\textbf{DAULAP} \cite{DBLP:conf/ijcai/ChengZY0LT019} is an anchor user prediction model based on the active learning method. It ensembles three query methods to estimate the most informative user pairs for the adaptive learning.

\item\textbf{MGCN} \cite{DBLP:conf/kdd/ChenYS0GM20} is a graph convolutional networks based alignment model. By defining various hypergraphs and integrating them into network embedding learning, it jointly learns representations for network vertices at different levels of granularity for the user alignment. 
\end{itemize}

We call the framework without the meta-learning as \underline{{\bf PS}}eudo anchor implanting (short for \underline{{\bf PS++}}). Besides, we call our framework as \underline{{\bf PS}}eudo anchor implanting based \underline{{\bf M}}eta \underline{{\bf L}}earning framework (short for {\bf PSML}) \footnote{The data and code are available in https://github.com/yanzihan1/PSML/tree/master}. Further, we want to evaluate the performance of different models with weights added to the anchors or with edges removed from the original network. We design \textbf{\{BaselineModel\}-AW} (Add Weight) as the baseline model running on networks whose weights of edges connected to anchors are doubled. Besides, \textbf{\{BaselineModel\}--} is the baseline model run on networks whose 5\% edges are randomly deleted.  

The hyper-parameters for the baseline models we used are reported in the original papers or the default settings of the open source codes provided by the authors. The learning rate of our PSML $\eta_1$ and $\eta_2$ are set as 0.01 and 0.0015 separately. For the evaluation metric, we use the 
$Precison@N$ \cite{DBLP:conf/ijcai/LiuCLL16,DBLP:conf/infocom/0002LZTWZ18,DBLP:journals/access/LiuZFZHZ19,DBLP:conf/kdd/ChenYS0GM20,DBLP:conf/ijcai/ChengZY0LT019}  for IONE, ABNE, DEEPLINK, DALUAP, and MGCN,  and $Hit-Precision$ \cite{DBLP:conf/aaai/LiWWYLLL19} for SNNA as these metrics are used in the original models.

\subsection{Strategies of Implanting Pseudo Anchors}
\begin{figure}[!ht]
\centering
\includegraphics[width=1\linewidth]{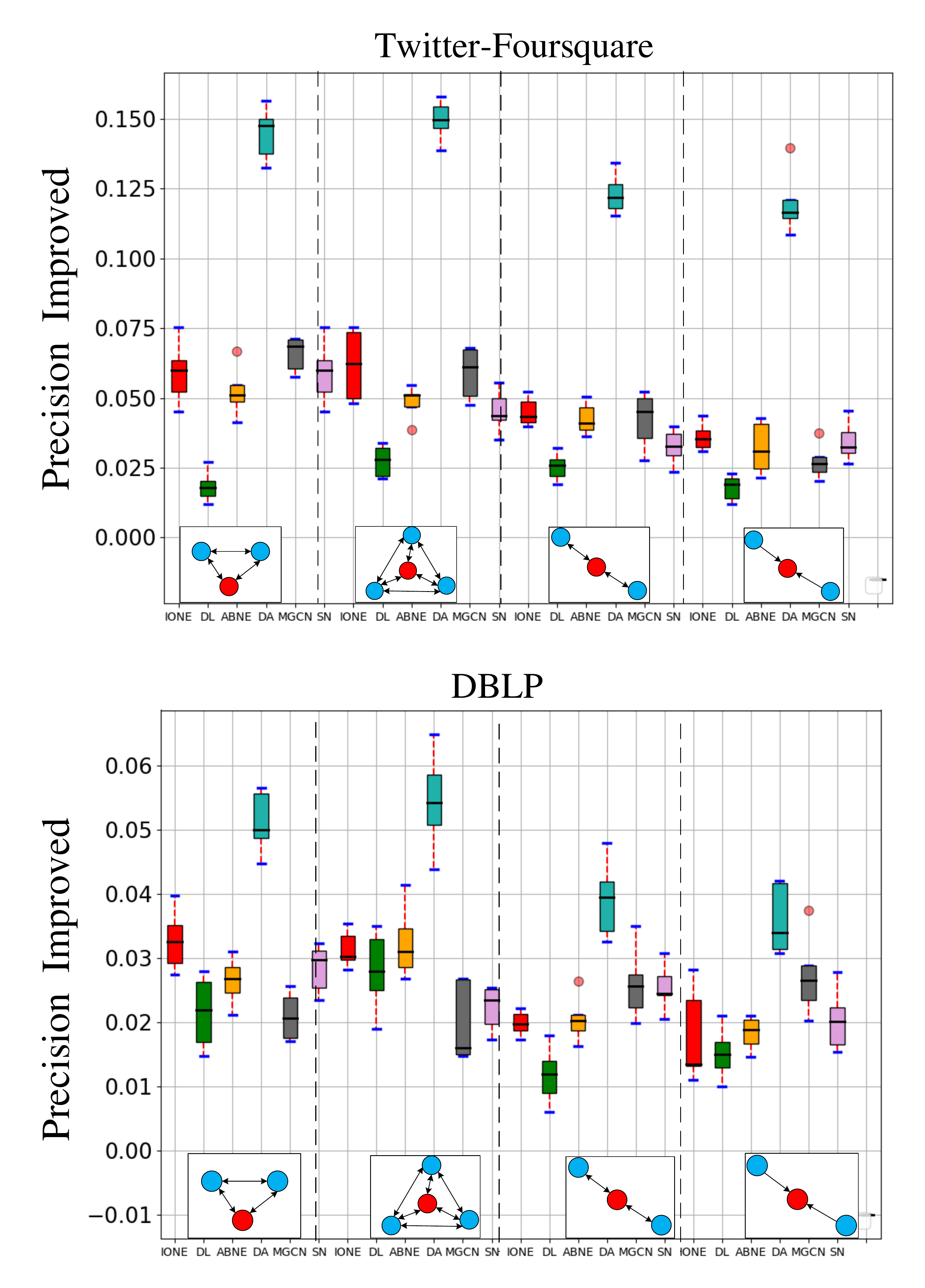}
\caption{Performance Evaluation Based on Different Strategies of Adding Pseudo Anchors }
\label{ExpPseudoAnchors}
\end{figure}

To determine the strategy of implanting the pseudo anchors, the factors to be considered include the number of pseudo anchors to be implanted (which can grow arbitrarily) and the connection patterns between the pseudo anchors and the real anchors (which can again increase exponentially as the number of pseudo anchors increased). For example, if the directed relationships among the pseudo and real anchors are to be considered, implanting two pseudo anchors for an real anchor will result in 36 different connection patterns \footnote{If bi-directional relationships are considered, the number of connection patterns among the pseudo anchors and the real anchor will be 3, and the number of connections between the pseudo anchors will be 4 as they can have no relationships. Based on this, the number of possible connection patterns is $3
\times 3 \times 4=36$}. For adding 3 pseudo anchors, 1,728 connection patterns will be resulted, which makes the enumeration of all possible patterns hard. We randomly choose some connection patterns as shown in Fig. \ref{ExpPseudoAnchors} and run different methods 5 times for each pattern. In Fig. \ref{ExpPseudoAnchors}, we illustrate also the box-plot of precision improvement compared with the original version of the baseline models based on the two datasets. In the figure, the bars in the column of each model denote the reasonable maximum and minimum values of 5 runs of experiments with the red dots being the outliers and the horizontal line in a rectangle showing the median value. From Fig. \ref{ExpPseudoAnchors}, we can see that implanting pseudo anchors is effective for enhancing the alignment quality, and the connection patterns with more edges between pseudo anchors and real anchors always give better performance. By balancing the complexity and efficiency of the implanting strategy, we choose the left-most pattern for the subsequent fine-tuning experiments.

\begin{figure*}[!htbp]
\includegraphics[width=0.95\linewidth]{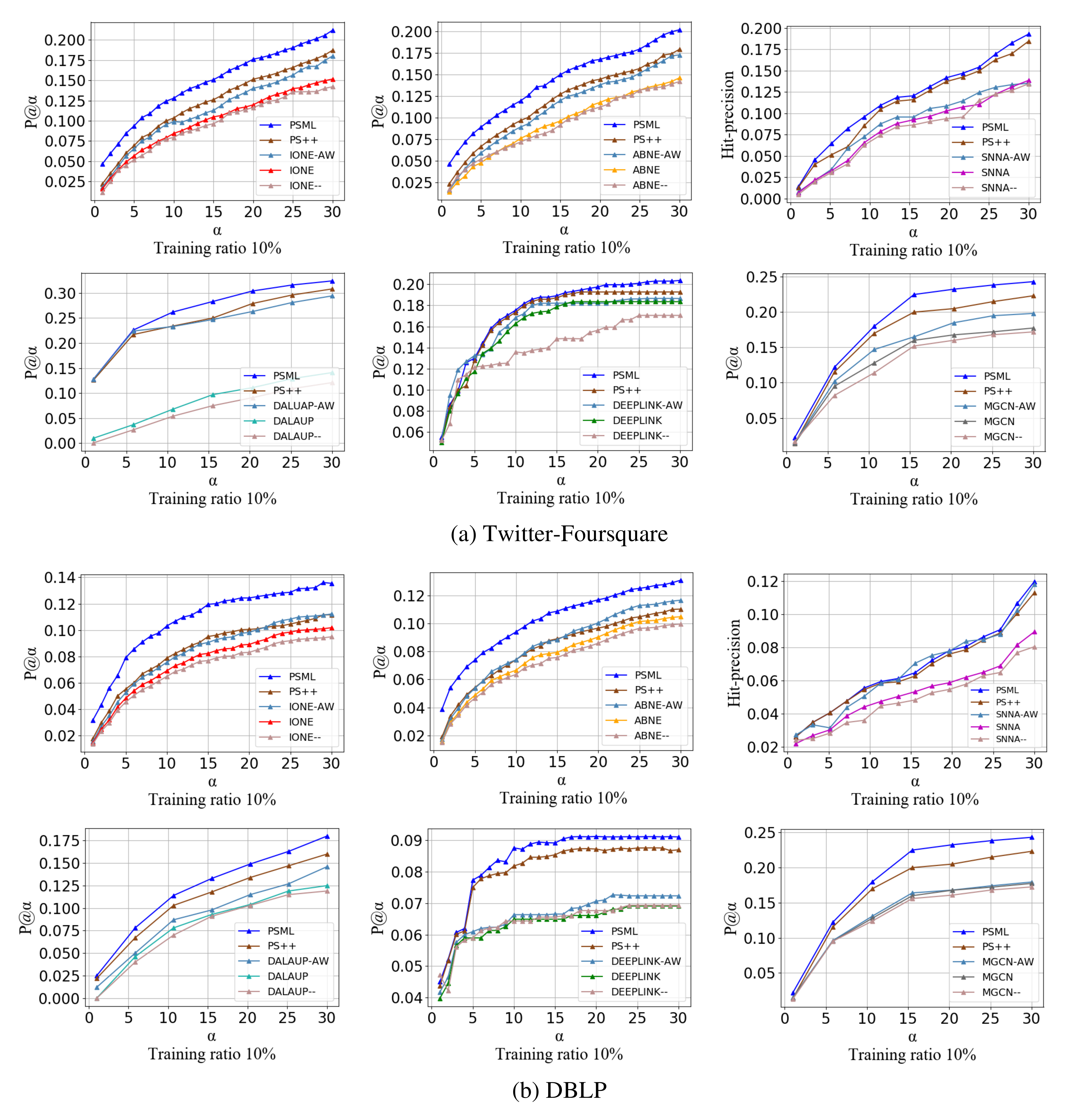} 
\caption{Performance Comparison between Proposed PSML and the Original Model}
\label{result_all}
\end{figure*}

\begin{figure}[!ht]
\centering
\includegraphics[width=1\linewidth]{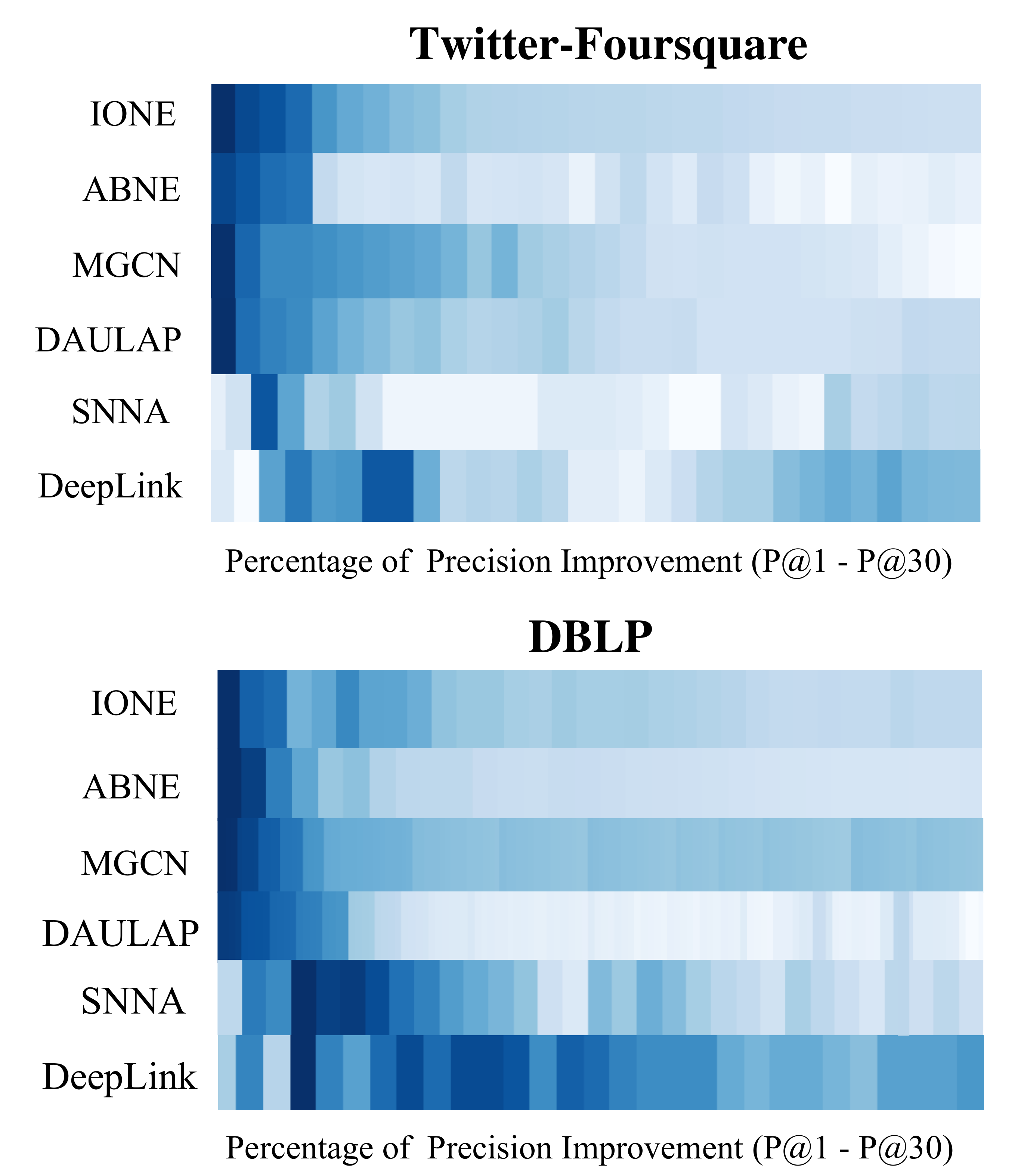}
\caption{Heat Map of PPI from $p@1$ to $p@30$}
\label{hot_fig}
\end{figure}

\subsection{Performance Analysis of Proposed Framework}
After determining the pattern of implanting pseudo anchors, we apply the fine-tuning strategy to the learning process for performance evaluation. In this section, we try to answer two questions: 1) whether the fine-tuning is useful for learning a better alignment space, 2) under what settings (such as $P@N$ and training ratio) our proposed framework can achieve the better improvement.

To answer the first question, we randomly select 10\% of anchors as training set and the rest as testing. For each comparison, we run for each model 5 times to obtain the average performance. According to Figure.\ref{result_all}.(a) and Figure.\ref{result_all}.(b), we can see that implanting pseudo anchors (PS++) shows superior performance compared with the original models under different settings of $P@N$  (Hit-precisions for SNNA) metrics, showing the effectness of implanting pseudo anchors. When the meta-learning algorithm is adopted (PSML), further improvement can be achieved when compared with the PS++ model. One possible reason is that, improper initial vectors of pseudo anchors may introduce negative influence in embedding learning as there is no fine-tuning of implanted anchors strategy in PS++. Although 5 times running and average values are used may reduce the influence of improper initial vectors for PS++ model, the PSML model still shows its superior performance in most of the case. Compared to the original version of baseline models, \{BaselineModel\}-AW (Add weight) achieves a better performance. We can conclude that \{BaselineModel\}-AW can learn a better alignment-oriented embedding space than the original baselines under the weakly supervised condition. The reason behind this is adding weights of edges that connect to anchors also can lead to first-order neighbors apart from others as anchors devote more influence to them. This is also consistent with our initial motivation. However, PS++ and PSML still have advantages compared to \{BaselineModel\}-AW, including controlling the connecting patterns and adjusting the positions based on meta-learning for learning a more evenly distributed embedding. Therefore, PSML and PS++ outperforms \{BaselineModel\}-AW in the most cases. Moreover, \{BaselineModel\}--, which runs the baseline model on the edge randomly deleted network, shows the worst performance compared to others. Since deleting edges will break the original characteristics of the original network, the model can not learn the faithful embedding for representing the original networks.


To answer the second question, 
we define the metric Percentage of Precision Improvement (PPI), 
given as
\begin{equation}
\label{PPI}
    PPI=\frac{Precision_{PSML}-Precision_{original}}{Precision_{original}}
\end{equation}
where $Precision_{PSML}$ and $Precision_{original}$ are the precision value of the proposed framework and that of the original model respectively.

Fig. \ref{hot_fig} shows the PPI of PSML at different $P@N$ settings where darker color shows higher precision improvement. In most cases, we can observe from Fig.\ref{hot_fig} that as the value of $N$ decreases, the PPI of PSML increases. When $N$ is small, our framework can lead to considerable improvement. A small value of $N$ means that we need to identify the potential anchors in a smaller search region. This implies more effort is required to avoid the ``fuzzy  regions'' as far as possible to ensure more accurate alignment. The ability to achieve this is due to the pulling effect of pseudo anchors and the fine-tuning process as explained.
In practice, high precision at small values of $N$ means we can effectively reduce the search space, which can in turn benefit downstream mechanisms such as co-training and active learning in many alignment models.

Furthermore, we evaluate the performance of the proposed framework in different training ratio settings, specifically from 3\%-15\%. Table \ref{train_ratio_imp} illustrates the results based on $P@30$. Our proposed framework shows better performance under different training ratio settings. Besides, in most cases, the PPI value increases with the decrease of the training ratio. PSML has superior performance especially when there is a relatively small number of training anchors. In practice, most potential anchors do not have direct relationships to anchors under extreme conditions of lacking labeled data. When they are in a dense local structure and without the supervision of anchors, there is a high probability to result in the ``overly-close'' phenomenon. While it will be hard for the original model to do the precise mapping across networks,
the PSML can achieve that by implanting pseudo anchors and controlling their update directions heuristically. This accounts for the improvement when there is a small number of anchors. Moreover, we observe that incorporating PSML into the DALUAP model has superior improvement compared to others. One reason for this is that the DALUAP is an active learning based algorithm. It labels potential anchors iteratively as the supervision for the next round. The mislabeling potential anchors in the previous steps
will result in accumulated errors in the subsequent steps. Incorporating PSML, to some extent, can alleviate the ``early errors'' in each epoch as it can learn the embedding space,  and thus result in a significant improvement.

\begin{table*}[!t]
\centering
\caption{Performance Comparison with Baseline Methods}
\begin{tabular}{l|llll| llll}
\hline
                        & \multicolumn{4}{c|}{Twitter-Foursquare}                           & \multicolumn{4}{c}{DBLP}                                                 \\ \hline
Training ratio        & 0.03           & 0.07           & 0.1            & 0.15           & 0.03                   & 0.07           & 0.1            & 0.15           \\ \hline
IONE                    & 5.24           & 11.02          & 15.01          & 18.77          & 3.34                   & 6.91           & 10.63          & 12.38          \\
PS++\_IONE              & 7.54           & 16.45          & 19.07          & 22.54          & 3.51                   & 7.54           & 11.56          & 13.55          \\
\textbf{PSML\_IONE}     & \textbf{9.67}  & \textbf{17.89} & \textbf{21.37} & \textbf{23.48} & \textbf{5.50}          & \textbf{9.50}  & \textbf{13.36} & \textbf{15.36} \\
PPI (Defined in Eq. \ref{PPI})                 & $\uparrow$ 84.54\%        & $\uparrow$ 62.34\%        & $\uparrow$ 42.37\%        & $\uparrow$ 25.09\%        & $\uparrow$ 64.67\%                & $\uparrow$ 37.48\%        & $\uparrow$ 25.68\%        & $\uparrow$ 24.07\%        \\ \hline
DEEPLINK                & 5.56           & 14.22          & 18.85          & 23.12          & 1.64                   & 3.64           & 6.94           & 9.00           \\
PS++\_DEEPLINK          & 10.14          & 16.00          & 19.77          & 24.31          & \textbf{2.76}          & 4.39           & 8.51           & 10.61          \\
\textbf{PSML\_DEEPLINK} & \textbf{10.14} & \textbf{16.17} & \textbf{21.37} & \textbf{24.50} & 2.75                   & \textbf{4.48}  & \textbf{9.14}  & \textbf{11.23} \\
PPI                   & $\uparrow$ 82.37\%        & $\uparrow$ 13.71\%        & $\uparrow$ 13.36\%        & $\uparrow$ 5.96\%         & $\uparrow$ 67.68\%                & $\uparrow$ 23.07\%        & $\uparrow$ 31.70\%        & $\uparrow$ 24.77\%        \\ \hline
ABNE                    & 5.54           & 9.64           & 14.91          & 18.98          & 3.23                   & 8.20           & 10.02          & 13.03          \\
PS++\_ABNE              & 7.31           & 12.41          & 17.45          & 21.87          & 3.43                   & 8.64           & 10.82          & 14.10          \\
\textbf{PSML\_ABNE}     & \textbf{9.23}  & \textbf{14.56} & \textbf{19.87} & \textbf{23.91} & \textbf{5.56}          & \textbf{10.88} & \textbf{13.72} & \textbf{16.45} \\
PPI                   & $\uparrow$ 66.60\%        & $\uparrow$ 51.03\%        & $\uparrow$ 33.26\%        & $\uparrow$ 25.97\%        & $\uparrow$ 72.13\%                & $\uparrow$ 32.68\%        & $\uparrow$ 36.92\%        & $\uparrow$ 26.24\%        \\ \hline
SNNA                    & 2.65           & 7.95           & 12.56          & 17.23          & 2.95                   & 4.66           & 8.65           & 9.89           \\
PS++\_SNNA              & 6.28           & 11.06          & 17.51          & 23.12          & 4.22                   & 6.56           & 11.74          & 13.20          \\
\textbf{PSML\_SNNA}     & \textbf{7.54}  & \textbf{11.96} & \textbf{18.09} & \textbf{24.04} & \textbf{4.32}          & \textbf{6.96}  & \textbf{11.91} & \textbf{13.89} \\
PPI                 & $\uparrow$ 184.52\%       & $\uparrow$ 50.44\%        & $\uparrow$ 44.02\%        & $\uparrow$ 39.52\%        & $\uparrow$ 46.44\%                & $\uparrow$ 49.35\%        & $\uparrow$ 37.68\%        & $\uparrow$ 40.44\%        \\ \hline
DALAUP                  & 4.37           & 8.76           & 14.12          & 19.92          & 2.45                   & 6.48           & 12.52          & 15.15          \\
PS++\_DALAUP            & 9.76           & 16.56          & 30.81          & 50.22          & 6.42                   & 15.67          & 16.02          & 20.42          \\
\textbf{PSML\_DALAUP}   & \textbf{10.45} & \textbf{17.89} & \textbf{33.38} & \textbf{53.24} & \textit{\textbf{7.83}} & \textbf{16.98} & \textbf{18.56} & \textbf{21.11} \\
PPI                   & $\uparrow$ 139.13\%       & $\uparrow$ 104.42\%       & $\uparrow$ 136.40\%       & $\uparrow$ 167.26\%       & $\uparrow$ 219.59\%               & $\uparrow$ 162.03\%       & $\uparrow$ 48.24\%        & $\uparrow$ 39.33\%        \\ \hline
MGCN                    & 5.23           & 14.62          & 17.95          & 21.13          & 2.45                   & 9.48           & 11.62          & 14.32          \\
PS++\_MGCN              & 7.98           & 18.75          & 22.32          & 27.68          & 3.96                   & 11.66          & 13.47          & 16.87          \\
\textbf{PSML\_MGCN}     & \textbf{9.45}  & \textbf{20.32} & \textbf{24.50} & \textbf{29.21} & \textbf{4.83}          & \textbf{11.98} & \textbf{14.56} & \textbf{17.11} \\
PPI                  & $\uparrow$ 80.68\%        & $\uparrow$ 38.98\%        & $\uparrow$ 36.49\%        & $\uparrow$ 38.23\%        & $\uparrow$ 97.14\%                & $\uparrow$ 26.37\%        & $\uparrow$ 25.30\%        & $\uparrow$ 19.48\%  \\ \hline
\end{tabular}
\label{train_ratio_imp}
\end{table*}

\begin{figure*}[ht]
\centering
\includegraphics[width=0.95\linewidth]{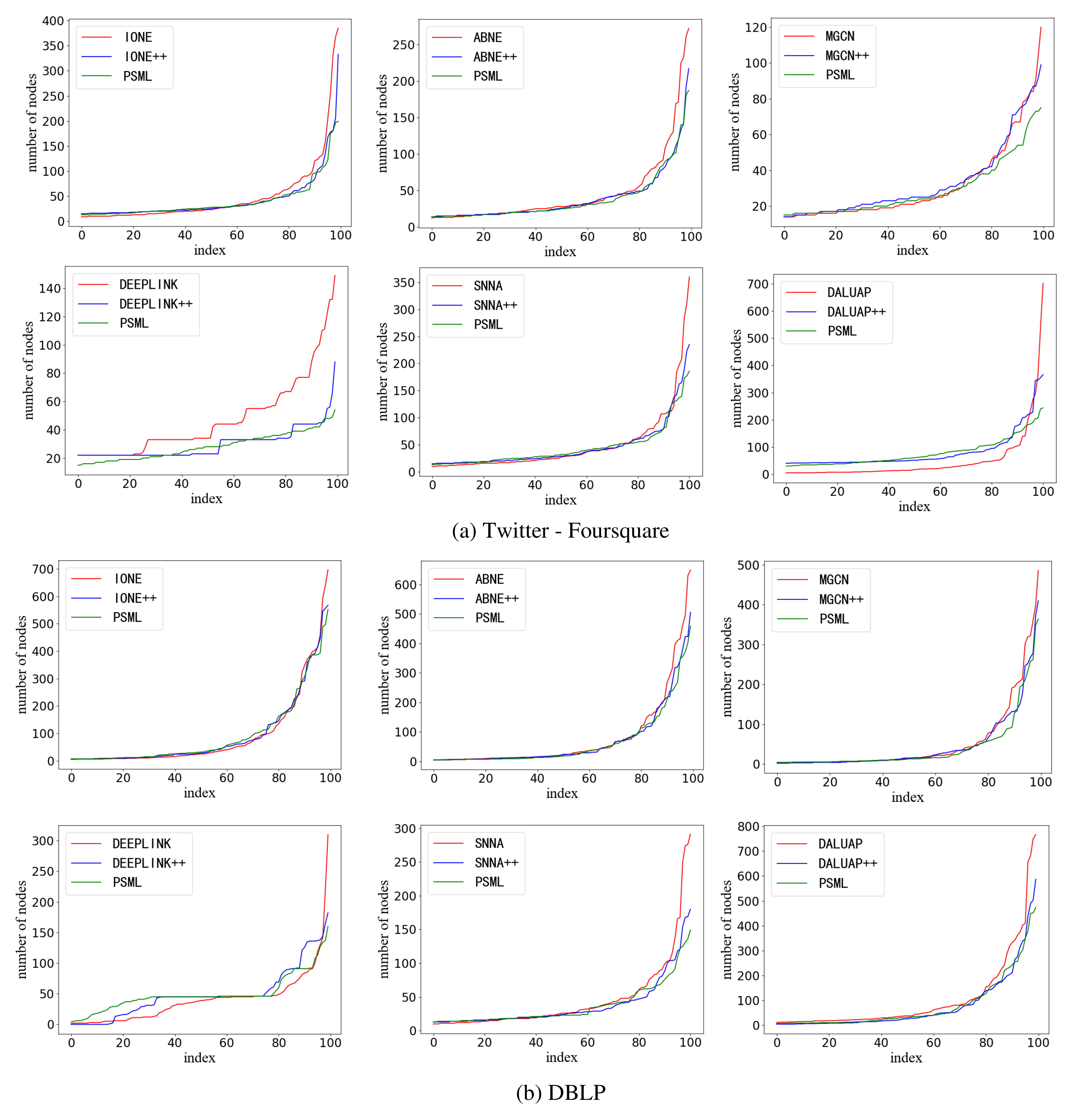}  
\caption{Distribution of Node Embeddings}
\label{EvenDistribution}
\end{figure*}

\subsection{Embedding Distribution}
We have explained that the improvement in precision is due to the capability to learn node embeddings which care more evenly distributed in the embedding space. In this section, we conduct experiments to illustrate this point empirically. Take the Foursquare network as an example. We first extract the embeddings learned by different models. We then apply the principal component analysis to project the embeddings on a 2-dimension space. Then, in order to estimate the density that can be comparable, we scale the learned embeddings to fit onto a $30\times{30}$ 2D plane (in other words, we use a $30\times{30}$ square of same size to cover all the embeddings) and then count how many users fall into each unit. 

We select the top 100 units with the most users for the analysis as many units have very few or no users. Fig. \ref{EvenDistribution} shows the frequency distribution of the number of users in the 100 unit. A frequency plot with a sharp growth implies that there are only a few units with a large number of users (indicating the ``overly-close'' phenomenon), while the others only have much less users. Meanwhile, a more flattened frequency plot denotes a more evenly distributed embedding space. In most cases, PSML can give more flattened distributions when compared with PS++ and the baselines. This indicates that our proposed framework can learn a more evenly distributed embedding space, which 
should in turn benefit the alignment task. This observation is consistent with the experimental results shown in Table \ref{train_ratio_imp} (Section 4.3). For DALUAP, IONE, and SNNA, the number of users that falls into the same unit can be as large as 700, 400, and 350 respectively. Incorporating PSML into these models shows great improvement in making the nodes to be much more evenly distributed in the embedding space. Therefore, compared with other baseline models, incorporating PSML into these models achieves a bigger improvement (139.12\% for DALUAP, 84.54\% for IONE, and 184.52\% for SNNA). We notice similar results based on the DBLP data. This implies that the improvement that a network alignment model can gain via the PSML is related to the embedding space learned by the original model. Our empirical results shows that models which learn less evenly distributed embedding can benefit more from PSML to achieve more significant performance improvement. This provides some clue to determine the applicable scope and the expected effect of the PSML framework.

\begin{figure}[!ht]
\centering
\includegraphics[width=0.8\linewidth]{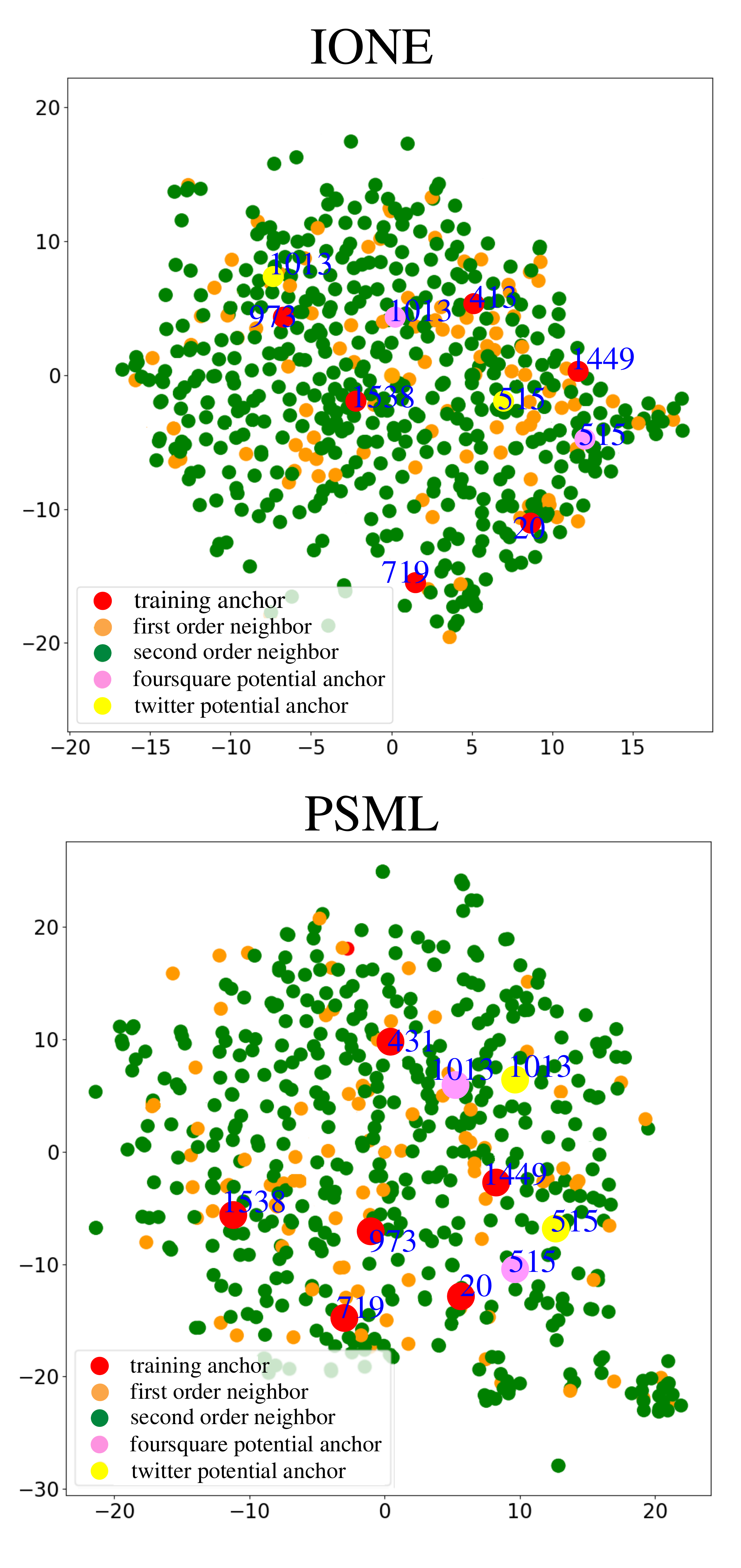}
\caption{Case Study (Macroscopic Level)}
\label{Studycase_1}
\end{figure}

\begin{figure}[ht]
\centering
\includegraphics[width=0.9\linewidth]{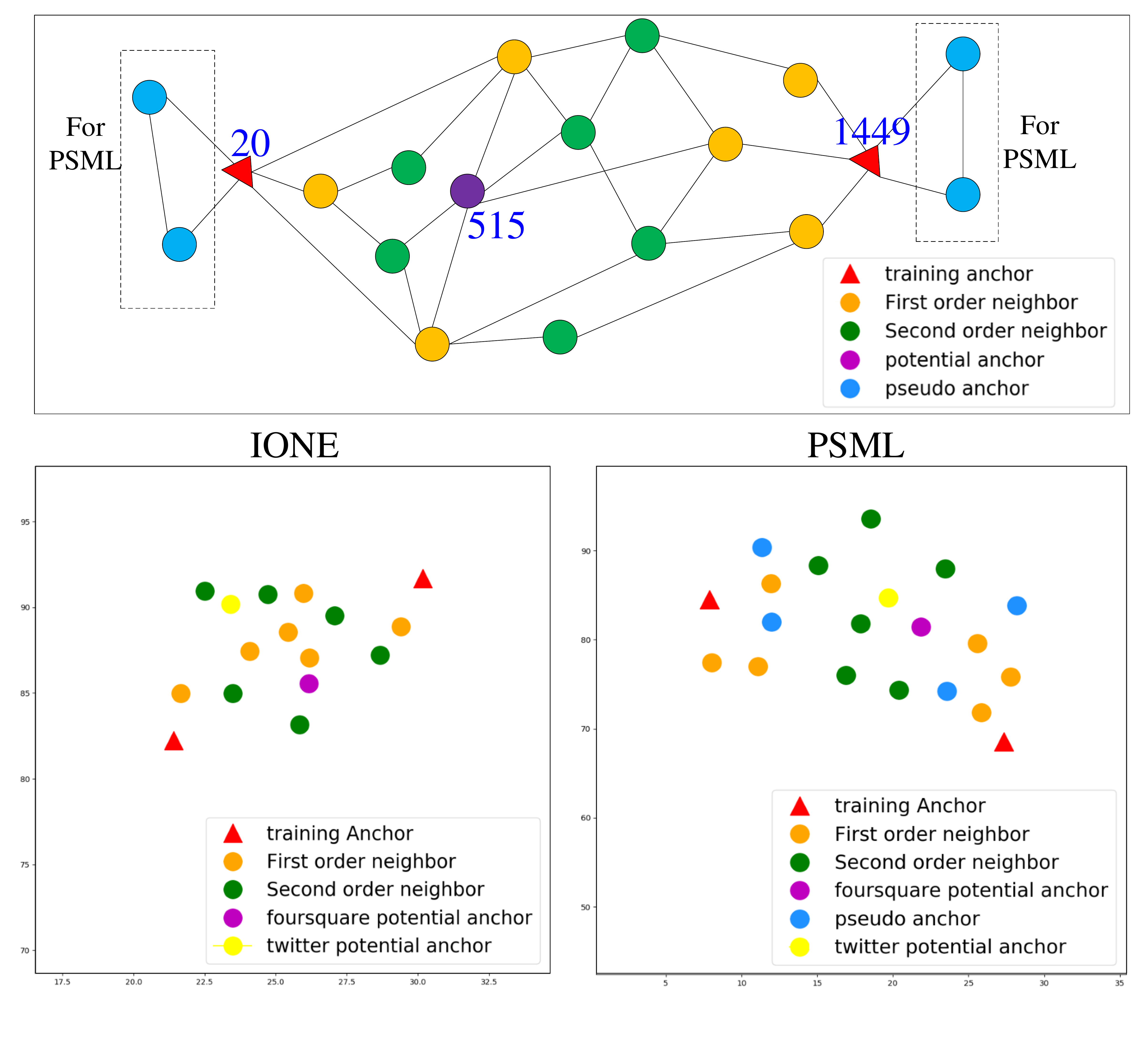}
\caption{Case Study (Microscopic Level)}
\label{Studycase}
\end{figure}

\subsection{Case Study}
Here we provide case studies to show the results obtained using the proposed framework is consistent with our conjecture. We first provide a case study to show at the macroscopic level that the PSML model can learn a more evenly distributed embedding space. Specifically, we select anchors whose ids are 719, 431, 1538, 1449, 20, and 973 
and their first-order and second-order neighbors in the Foursquare network (there are 560 nodes in total).  Then we extract the corresponding embeddings learned by IONE and PSML. Fig. \ref{Studycase_1} shows the low-dimensional representations of these embeddings obtained by t-SNE \cite{tsne} respectively. Red dots represent the anchors, while orange and green dots represent first-order and second-order neighbors of the anchors. Dots in purple are potential anchors, and we also plot dots in yellow to indicate the corresponding anchors in twitter.
In Fig. \ref{Studycase_1}, it is obvious that PSML can learn a more even distributed embedding space compared with the original IONE model. Nodes around potential anchors (dots in purple and yellow) are evenly distributed. This embedding space can thus benefit the subsequent mapping operations for alignment.

We then take a more microscopic view and provide a case study where we select only a sub-network formed by anchors 20 and 1449 and their neighbors in the Twitter-Foursquare dataset. Then, we use t-SNE to reduce the dimension of embedding for visualization. Fig. \ref{Studycase} illustrates the embeddings obtained by IONE and PSML separately. The red triangles represent two anchor nodes, the orange dots represent the first-order neighbors of the anchor node, the green dots represent the second-order neighbors of the anchor nodes, and the dots in purple and yellow colors represent the potential anchors across the networks. According to Fig. \ref{Studycase}, we have the following observations. Without the pseudo anchors implanted, the first-order neighbors of the anchors in the IONE model are hard to be distinguished from others as they are in an ``overly-close'' region. For PSML, the first-order and the second-order neighbors are more properly organized so that the former ones are closer to the anchors as compared with the latter ones. This is consistent to the desirable organization of the embeddings.


\section{Conclusion}
This paper studies how to improve the embedding based alignment model across social networks via pseudo anchors. We implant pseudo anchors to each labeled anchors and develop a corresponding meta-learning algorithm to fine-tune the embedding of the pseudo anchors for better alignment performance. The proposed framework named PSML can be integrate into most of the existing embedding based network alignment models to learn a more evenly distributed embedding space across networks for enhancing the alignment accuracy. 
By integrating PSML into several state-of-the-art network alignment models, our experimental results demonstrate that our framework can successfully enhance the performance of many STOA embedding based models. Future research directions include further optimizing PSML with automatic determination of the number of pseudo anchors and the corresponding connecting patterns.


%

\section*{Acknowledgments}

The work is partially supported by National Natural Science Foundation of China (61936001, 61806031), and in part by the Natural Science Foundation of Chongqing (cstc2019jcyj-cxttX0002), and in part by the key cooperation project of Chongqing  Municipal Education Commission (HZ2021008), and in part by Doctoral Innovation Talent Program of Chongqing University of Posts and Telecommunications (BYJS202118). This work is partially done when Li Liu works at Hong Kong Baptist University supported by the Hong Kong Scholars program (XJ2020054).

\ifCLASSOPTIONcaptionsoff
  \newpage
\fi



%

\bibliographystyle{IEEEtran}
\bibliography{IEEEabrv,aligment}




%

\begin{IEEEbiography}[{\includegraphics[width=1in,height=1.25in,clip,keepaspectratio]{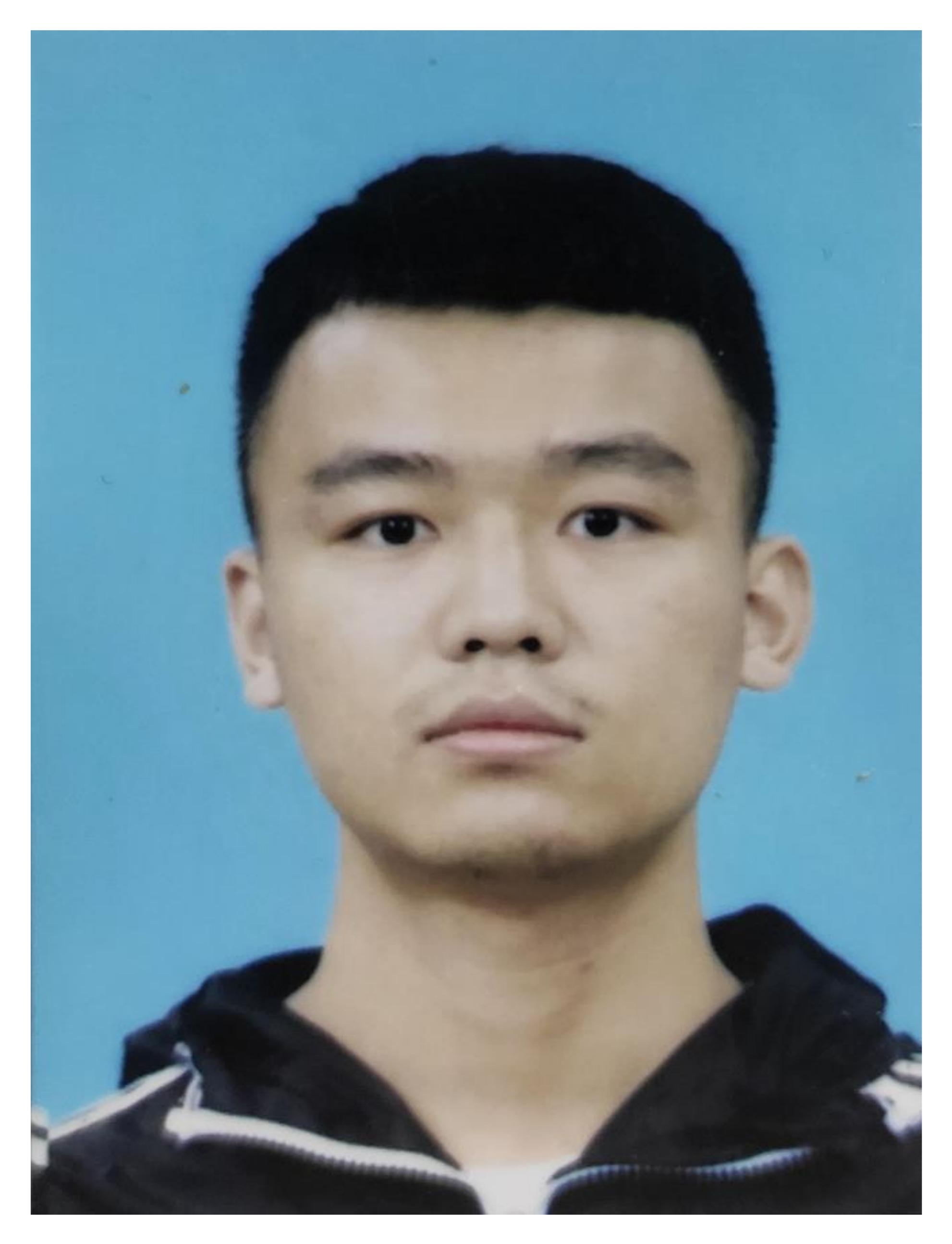}}]{Zihan Yan}
received his B.S. degree in Computer
science from the Chongqing Institute of Engineering, China, in 2015. He is currently pursuing a M.S.
degree in computer technology at the Chongqing
University of Posts and Telecommunications in
Chongqing, China. His research interests include
social network and data mining.
\end{IEEEbiography}

\begin{IEEEbiography}[{\includegraphics[width=1in,height=1.25in,clip,keepaspectratio]{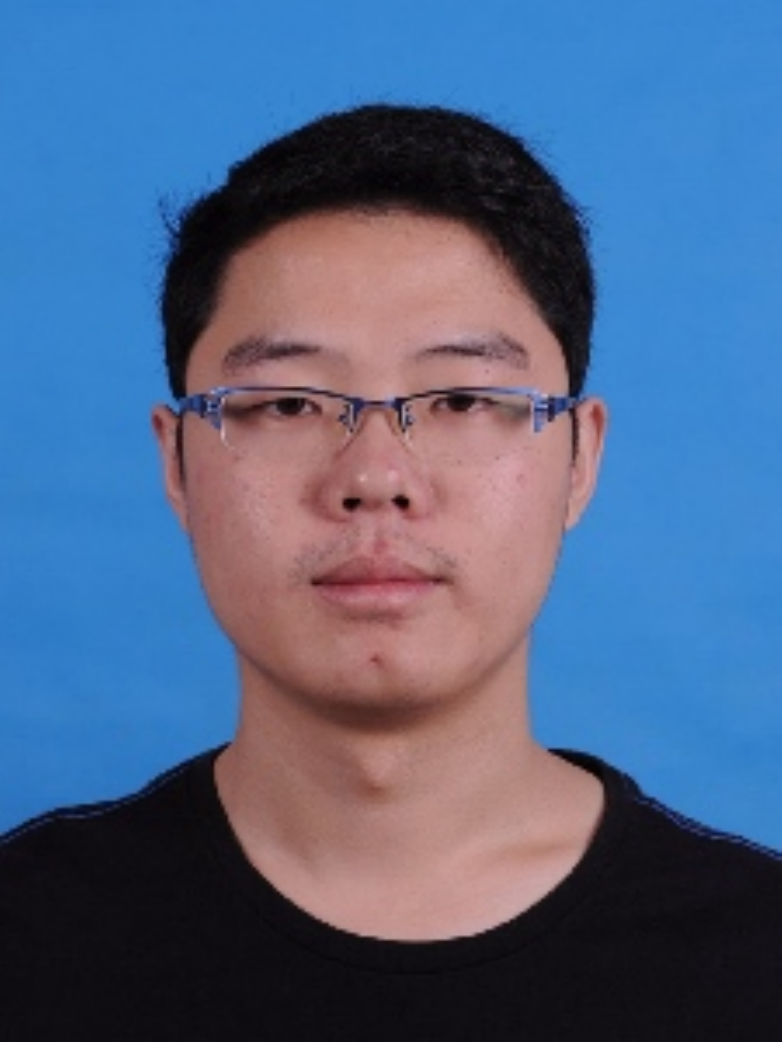}}]{Li Liu}
received the  Ph.D. in computer science from Beijing Institute of Technology in 2016, the M.E. in computer science from Kunming University of Science and Technology in 2012 and the B.B.A. in Information management and information system from Chongqing University of Posts and Telecommunications in 2009. He was a visiting student at Hong Kong Baptist University during 2016. He is currently an Associate Professor in Chongqing University of Posts and Telecommunications. His research interests include web mining and social computing.
\end{IEEEbiography}


\begin{IEEEbiography}[{\includegraphics[width=1in,height=1.25in,clip,keepaspectratio]{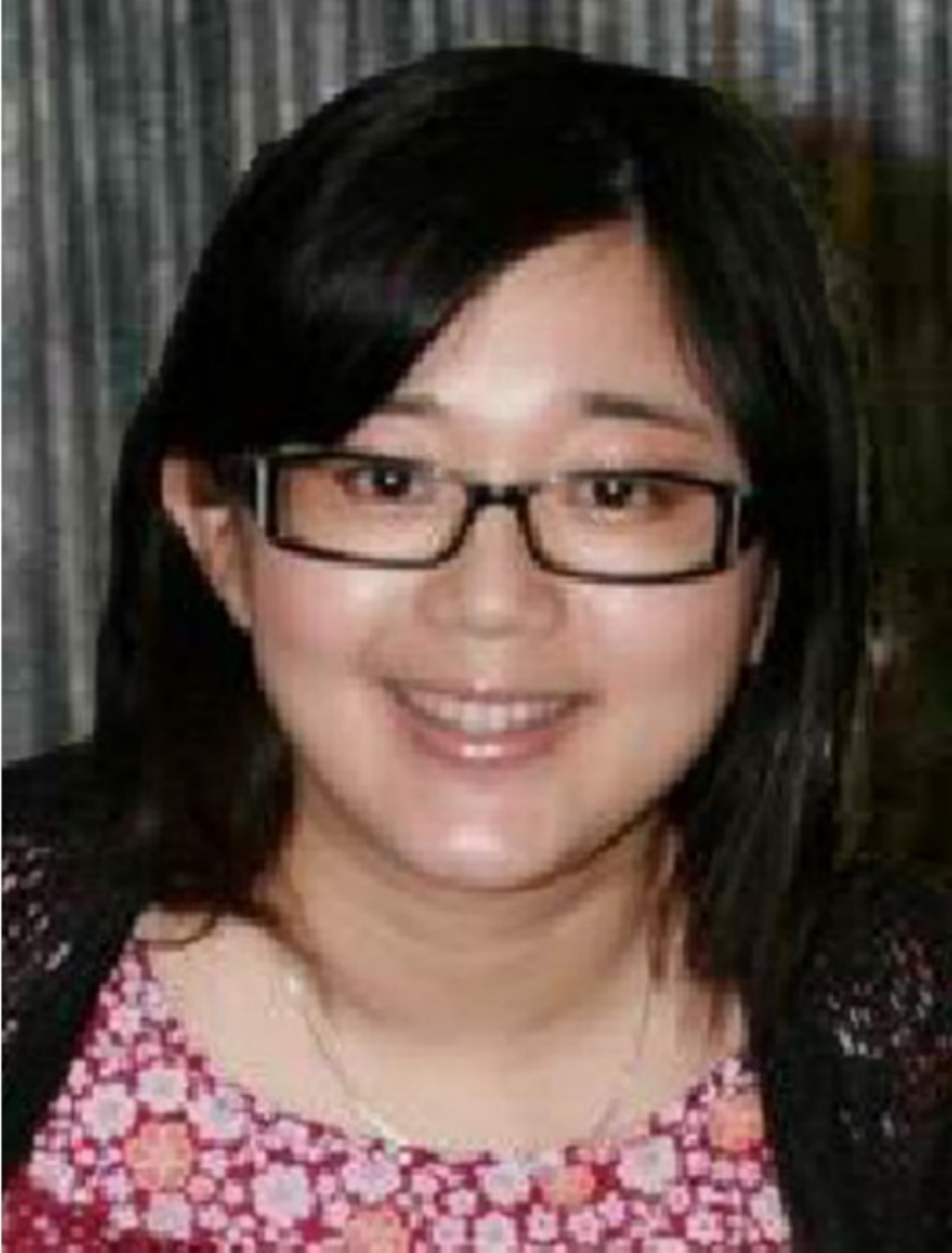}}]{Xin Li}
is currently an Associate Professor in the School of Computer Science at Beijing Institute of Technology, China.  She received the B.Sc. and M.Sc degrees in Computer Science from Jilin University  China, and the Ph.D. degree in Computer Science at Hong Kong Baptist University. Her research focuses on the development of algorithms for representation learning, reasoning under uncertainty and machine learning with application to Natural Language Processing, Recommender Systems, and Robotics.
\end{IEEEbiography}

\begin{IEEEbiography}[{\includegraphics[width=1in,height=1.25in,clip,keepaspectratio]{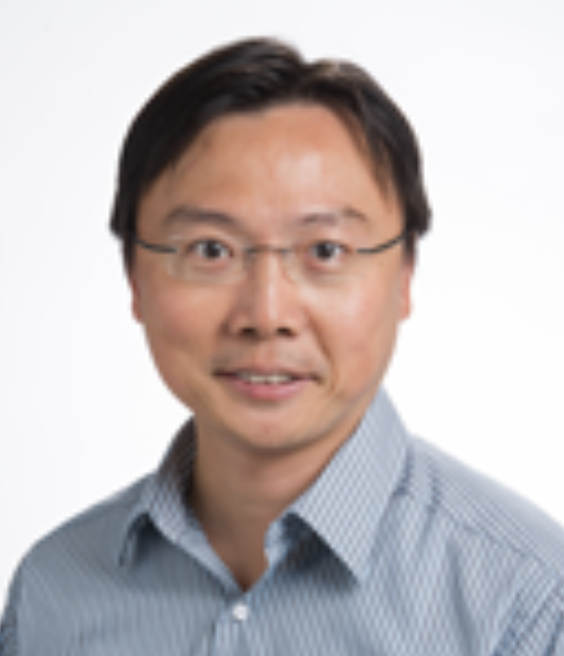}}]{William K. Cheung} received the Ph.D. degree in computer science from the Hong Kong University of Science and Technology in Hong Kong in 1999. He is currently Professor of the Department of Computer Science, Hong Kong Baptist University, Hong Kong. His current research interests include artificial intelligence, data mining, collaborative information filtering, social network analysis, and healthcare informatics. He has served as the Co-Chairs and Program Committee Members for a number of international conferences and workshops, as well as Guest Editors of journals on areas including artificial intelligence, Web intelligence, data mining, Web services, e-commerce technologies, and health informatics. From 2002-2018, he was on the Editorial Board of the IEEE Intelligent Informatics Bulletin. He is currently a Track Editor of Web Intelligence Journal and an Associate Editor of Journal of Health Information Research, and Network Modeling and Analysis for Health Informatics and Bioinformatics.
\end{IEEEbiography}

\begin{IEEEbiography}[{\includegraphics[width=1in,height=1.25in,clip,keepaspectratio]{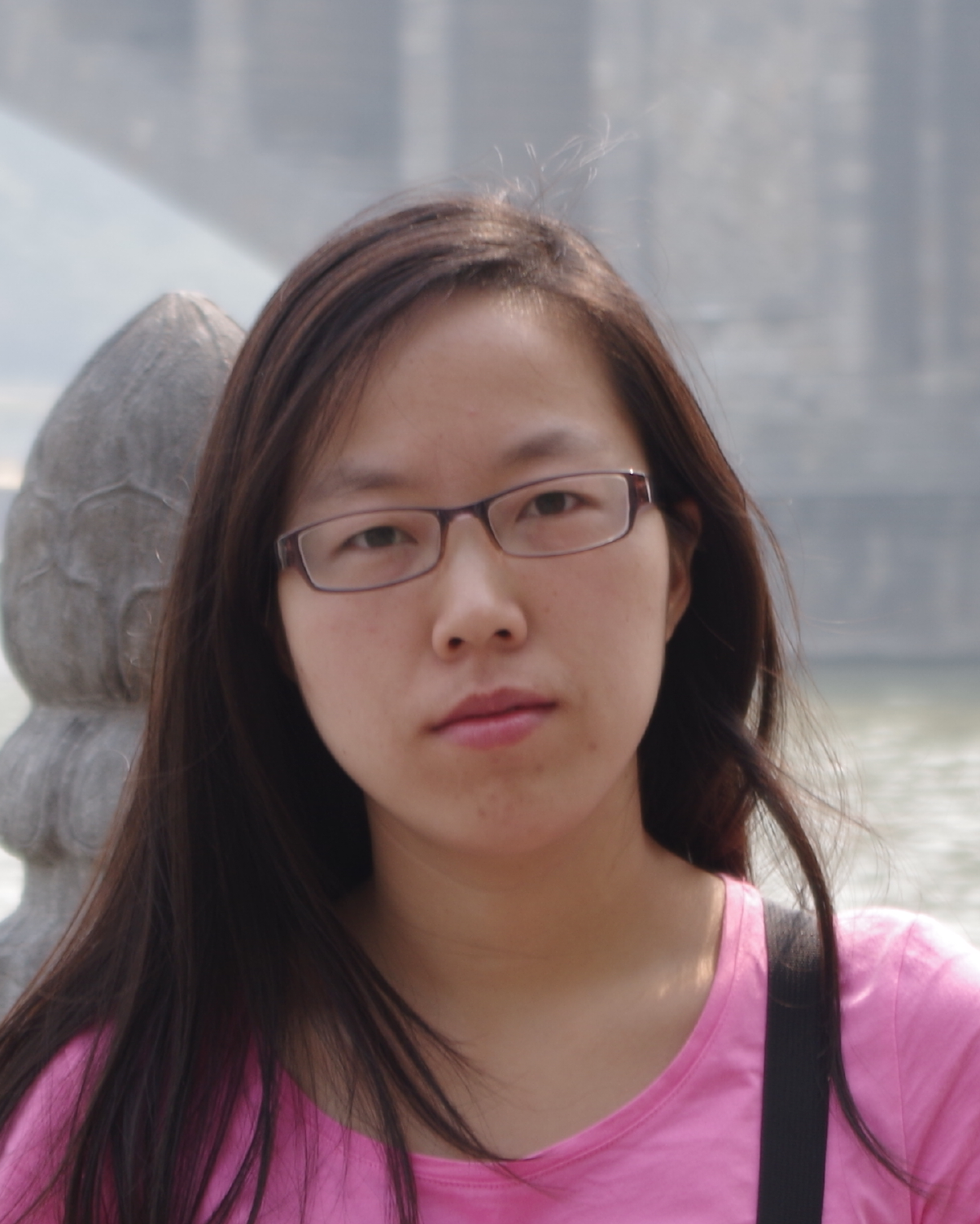}}]
{Youmin Zhang} received M.E. in control engineering from Kunming University of Science and Technology in 2013 and the B.E. in automation from University of Jinan 2009. She is currently pursuing a Ph.D. degree in computer technology at the Chongqing
University of Posts and Telecommunications in
Chongqing, China. She is currently an Assistant Professor in Chongqing Institute of Engineering. Her research interests include social computing and knowledge graph mining.
\end{IEEEbiography}

\begin{IEEEbiography}[{\includegraphics[width=1in,height=1.25in,clip,keepaspectratio]{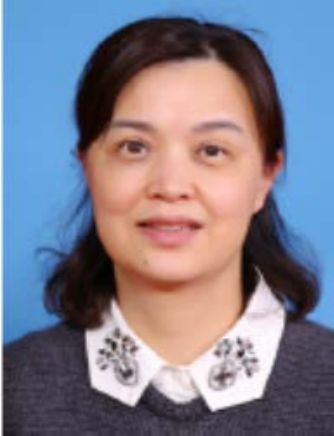}}]{Qun Liu} received her B.S. degree from Xi’An Jiaotong University in China in 1991, and the M.S. degree from Wuhan University in China in 2002, and the Ph.D from Chongqing University in China in 2008. She is currently a Professor with Chongqing University of Posts and Telecommunications. Her current research interests include complex and intelligent systems, neural networks and intelligent information processing.
\end{IEEEbiography}

\begin{IEEEbiography}[{\includegraphics[width=1in,height=1.25in,clip,keepaspectratio]{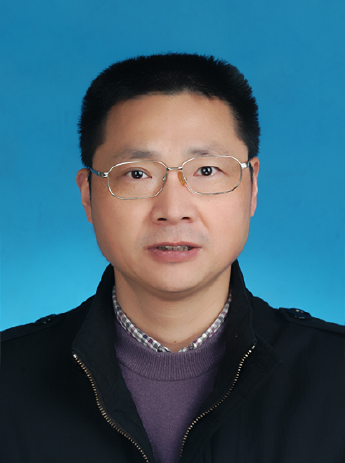}}]{Guoyin Wang}(SM’03) received the B.S., M.S., and Ph.D. degrees from Xi’an Jiaotong University, Xian, China, in 1992, 1994, and 1996, respectively. He was at the University of North Texas, and the University of Regina, Canada, as a visiting scholar during 1998-1999. Since 1996, he has been at the Chongqing University of Posts and Telecommunications, where he is currently a professor, the director of the Chongqing Key Laboratory of Computational Intelligence, the Vice-President of the University and the dean of the School of Graduate. He was appointed as the director of the Institute of Electronic Information Technology, Chongqing Institute of Green and Intelligent Technology, CAS, China, in 2011. He is the author of over 10 books, the editor of dozens of proceedings of international and national conferences, and has more than 300 reviewed research publications. His research interests include rough sets, granular computing, knowledge technology, data mining, neural network, and cognitive computing, etc. Dr. Wang was the President of International Rough Set Society (IRSS) 2014-2017. He is a Vice-President of the Chinese Association for Artificial Intelligence (CAAI), and a council member of the China Computer Federation (CCF).

\end{IEEEbiography}
\end{document}


%
\title{Towards Improving Embedding Based Models of Social Network Alignment via Pseudo Anchors Appendices}
%
%
%
%

\author{Zihan~Yan,~
        Li~Liu,~
        Xin~Li,~\IEEEmembership{Member,~IEEE},
        William K.~Cheung,~\IEEEmembership{Member,~IEEE},
        Youmin~Zhang,~
        Qun~Liu,~
        and ~Guoyin~Wang,~\IEEEmembership{Senior Member,~IEEE}}
\markboth{Journal of \LaTeX\ Class Files,~Vol.~14, No.~8, August~2015}%
{Shell \MakeLowercase{\textit{et al.}}: Bare Demo of IEEEtran.cls for Computer Society Journals}
%



\maketitle
\appendices
\section{Proof of ``The order of proximity increases, the amount of shifting will decrease accordingly"}

Here, we provide proof we describe in subsection 3.1.  With the relationship orders increase, the shifting will decrease accordingly.

According to Eq. (7) and (8) in subsection 3.1, the shiftings of $v_b$ and $v_c$ are shown in Eq. (\ref{shiftb}) and  Eq. (\ref{shiftc}).
\begin{equation} \label{shiftb}
     \Delta{\overrightarrow{u}_b} \propto \sigma({\overrightarrow {u}_b}^T (\overrightarrow {u_{a}}+\Delta{\overrightarrow{u}_a}))-\sigma({\overrightarrow {u}_b}^T \overrightarrow {u_{a}})
\end{equation}

\begin{equation} \label{shiftc}
    \Delta{\overrightarrow{u}_c} \propto \sigma({\overrightarrow {u}_c}^T (\overrightarrow {u_{b}}+\Delta{\overrightarrow{u}_b}))-\sigma({\overrightarrow {u}_c}^T \overrightarrow {u_{b}})
\end{equation}

We firstly consider the shifting of $v_b$ as a function of $\overrightarrow{u_a}$ as it is affected by node $v_a$.
\begin{equation} \label{shiftb_fun}
    f(\overrightarrow{u_a})= \sigma({\overrightarrow {u}_b}^T (\overrightarrow {u_{a}}+\Delta{\overrightarrow{u}_a}))-\sigma({\overrightarrow {u}_b}^T \overrightarrow {u_{a}})
\end{equation}

Then we use Taylor Series to decompose first part of $f(\overrightarrow{u_a})$, $\sigma({\overrightarrow {u}_b}^T (\overrightarrow {u_{a}}+\Delta{\overrightarrow{u}_a}))$, which is shown in Eq. (4)


\begin{equation*} \label{shiftb_taylor}
    \sigma({\overrightarrow {u}_b}^T (\overrightarrow {u_{a}}+\Delta{\overrightarrow{u}_a}))= \sigma({\overrightarrow {u}_b}^T \overrightarrow {u_{a}})+\sigma^{'}({\overrightarrow {u}_b}^T \overrightarrow {u_{a}})\Delta{\overrightarrow{u}_a}+\frac{\sigma^{''}({\overrightarrow {u}_b}^T \overrightarrow {u_{a}})}{2!}\Delta{\overrightarrow{u}_a^2}+...+\frac{\sigma^{(n)}({\overrightarrow {u}_b}^T \overrightarrow {u_{a}})}{n!}\Delta{\overrightarrow{u}_a^n}+...
    \eqno{(4)}
\end{equation*}

Then   $\Delta{\overrightarrow{u}_b}$ can be written as follows.

\begin{eqnarray*}\label{re_write_deltab}
    \Delta{\overrightarrow{u}_b}&=& \sigma^{'}({\overrightarrow {u}_b}^T \overrightarrow {u_{a}})\Delta{\overrightarrow{u}_a}+\frac{\sigma^{''}({\overrightarrow {u}_b}^T \overrightarrow {u_{a}})}{2!}\Delta{\overrightarrow{u}_a^2}+...+\frac{\sigma^{(n)}({\overrightarrow {u}_b}^T \overrightarrow {u_{a}})}{n!}\Delta{\overrightarrow{u}_a^n}+... \\
    &=&\Delta{\overrightarrow{u}_a}(\sigma^{'}({\overrightarrow {u}_b}^T \overrightarrow {u_{a}})+\frac{\sigma^{''}({\overrightarrow {u}_b}^T \overrightarrow {u_{a}})}{2!}\Delta{\overrightarrow{u}_a}+...+\frac{\sigma^{(n)}({\overrightarrow {u}_b}^T \overrightarrow {u_{a}})}{n!}\Delta{\overrightarrow{u}_a^{n-1}}+...)\\
    &<&\Delta{\overrightarrow{u}_a}(\sigma^{'}({\overrightarrow {u}_b}^T \overrightarrow {u_{a}})+\frac{(\sigma^{'})^{'}({\overrightarrow{u}_b}^T \overrightarrow {u_{a}})}{1!}\Delta{\overrightarrow{u}_a}+...+\frac{\sigma^{'(n-1)}({\overrightarrow {u}_b}^T \overrightarrow {u_{a}})}{n-1!}\Delta{\overrightarrow{u}_a^{n-1}}+...)\\
    &=& \Delta{\overrightarrow{u}_a}(\sigma'({\overrightarrow {u}_b}^T (\overrightarrow {u_{a}}+\Delta{\overrightarrow{u}_a}))) 
\end{eqnarray*}

Since $\sigma '=(1-\sigma)\times{\sigma}<1$, then $\Delta{\overrightarrow{u}_b} < \Delta{\overrightarrow{u}_a}$. Note here we use the sigmoid function for similarity calculation as it is widely adopted for scaling the absolute similarity into a unified range (0,1). The relative similarity in range (0,1) is the main factor that reflects the distribution of embedding space. We argue that other functions which reflect the relative similarity can give the same result.

Similarly, we use Taylor Series to decompose $\sigma({\overrightarrow {u}_c}^T (\overrightarrow {u_{c}}+\Delta{\overrightarrow{u}_b}))$. It is shown in Eq. (5).

\begin{equation*} \label{shiftc_taylor}
    \sigma({\overrightarrow {u}_c}^T (\overrightarrow {u_{c}}+\Delta{\overrightarrow{u}_b}))= \sigma({\overrightarrow {u}_c}^T \overrightarrow {u_{b}})+\sigma^{'}({\overrightarrow {u}_c}^T \overrightarrow {u_{b}})\Delta{\overrightarrow{u}_b}+\frac{\sigma^{''}({\overrightarrow {u}_c}^T \overrightarrow {u_{b}})}{2!}\Delta{\overrightarrow{u}_b^2}+...+\frac{\sigma^{(n)}({\overrightarrow {u}_c}^T \overrightarrow {u_{b}})}{n!}\Delta{\overrightarrow{u}_b^n}+...
     \eqno{(5)}
\end{equation*}

Then the we $\Delta{\overrightarrow{u}_c}$ as Eq. (6).

\begin{equation*} \label{re-write_c}
    \Delta{\overrightarrow{u}_c}= \sigma^{'}({\overrightarrow {u}_c}^T \overrightarrow {u_{b}})\Delta{\overrightarrow{u}_b}+\frac{\sigma^{''}({\overrightarrow {u}_c}^T \overrightarrow {u_{b}})}{2!}\Delta{\overrightarrow{u}_b^2}+...+\frac{\sigma^{(n)}({\overrightarrow {u}_c}^T \overrightarrow {u_{b}})}{n!}\Delta{\overrightarrow{u}_b^n}+...=\Delta{\overrightarrow{u}_b}(\sigma'({\overrightarrow {u}_c}^T (\overrightarrow {u_{b}}+\Delta{\overrightarrow{u}_b})))
    \eqno{(6)}
\end{equation*}

Since $\sigma '=(1-\sigma)\times{\sigma}<1$, $\Delta{\overrightarrow{u}_c} < \Delta{\overrightarrow{u}_b}$. Therefore, we obtain $\Delta{\overrightarrow{u}_c} < \Delta{\overrightarrow{u}_b} < \Delta{\overrightarrow{u}_a}$. This completes the proof.

\clearpage
\section{Incorporating PSML into Hyperbolic Space based Alignment Model}

To investigate the user alignment in hyperbolic space, several works conduct the alignment in the hyperbolic space for better fitting to the power-law distribution and hierarchical structures in the network data [45], [46]. For instance, the distance between nodes can be defined using the Lorentz model [45] and Poincar{é} ball model [46]. The random walk-based structural proximity objectives can then be optimized for the alignment.

To see the performance of incorporating PSML into a hyperbolic embedding based model. We adopt the HUIL [45] that uses the Lorentz model in hyperbolic space as the baseline model. It defines a Riemannian manifold shown as Eq. (\ref{Riemannian}).

\begin{equation} \label{Riemannian}
    \mathbf{L}^{n}=\left\{x\in R^{n+1} |x^2_{0}-\sum\limits_{i=1}^{n}x^2_{i}=1 \bigwedge x_{0}>0 \right  \}
\end{equation}

Further, the distance metric between vectors is defined as Eq.(\ref{distance}):
\begin{eqnarray} \label{distance}
    \nonumber
    d_{\mathbf{L}\left(x,y\right)}=arcosh\left(-\left\langle{x,y}\right\rangle_{\mathbf{L}}\right)\\     
    x,y\in R^{n+1}, 
    \left\langle{x,y}\right\rangle_{\mathbf{L}}=-x_{0}y_{0}+\sum\limits_{i=1}^{n}x_{i}y_{i}
\end{eqnarray}

By the definition of Eq. (\ref{Riemannian}) and Eq. (\ref{distance}), HULL utilizes a shallow neural network to construct the structure preserving objective function. In addition, we conduct the same experiment described in Subsection 4.4.

\begin{figure*}[!htbp]
\includegraphics[width=1\linewidth]{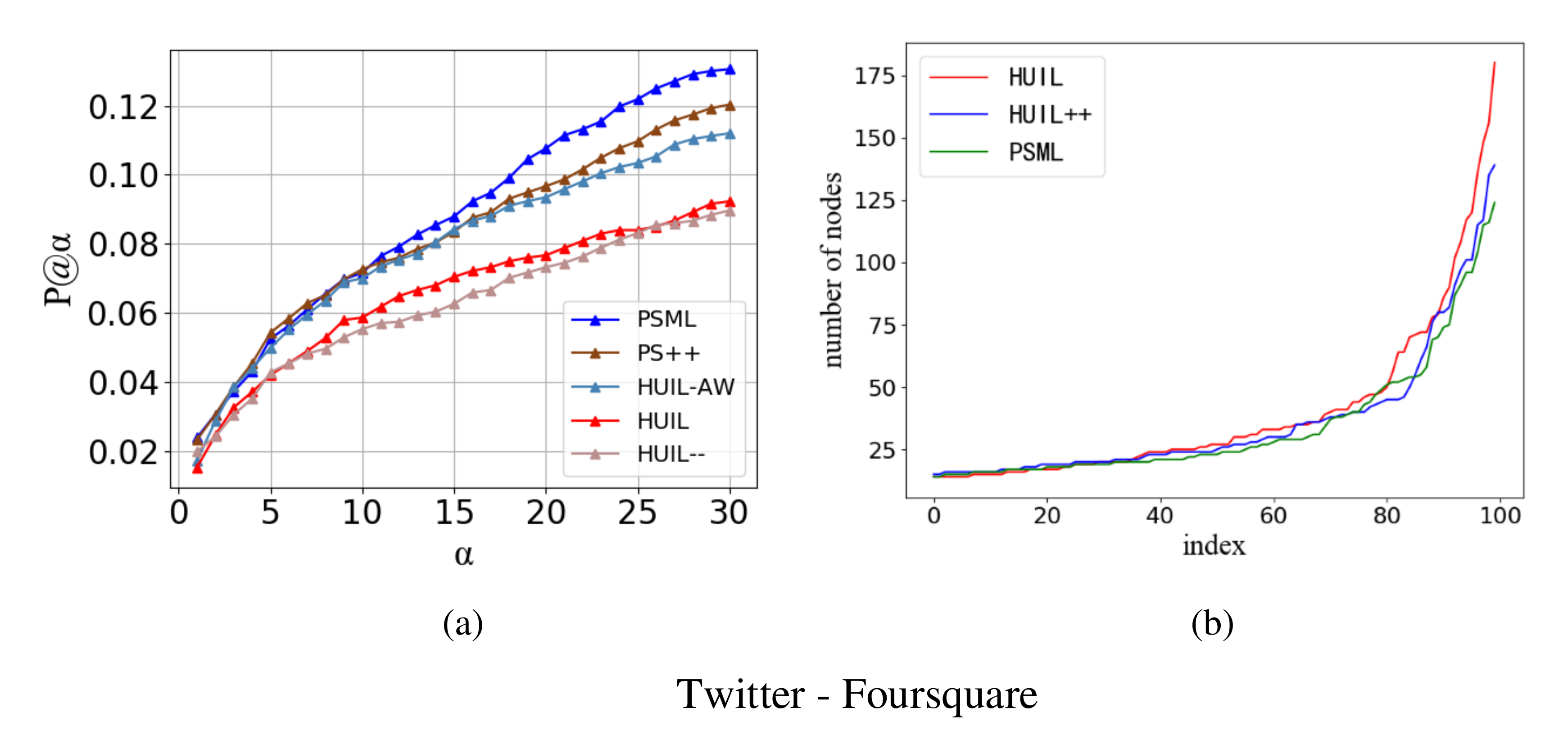} 
\caption{Performance of Incorporating PSML into Hyperbolic Space Based Alignment Model}%
\label{result_hyperbolic}
\end{figure*}

The experimental results of Twitter-Foursquare are shown in Fig. \ref{result_hyperbolic}. From Fig. \ref{result_hyperbolic}. (a), we observe that incorporating PSML into the hyperbolic embedding model benefits the performance in precision. Moreover, we also conduct the experiment described in Subsection 4.5.  In Fig. \ref{result_hyperbolic}. (b),  we observe that the PSML framework can also learn a more even distribution compared to the original one. Note it is the very primary study on embedding distribution on hyperbolic space that is far from a compact experimental analysis, we place this analysis in the Appendix of this paper. Hope it can provide some inspiration for the following studies.

